\documentclass[letterpaper,twocolumn,10pt]{article} 
\usepackage{usenix2019_v3} 
\usepackage{cite}
\usepackage{amsmath,amssymb,amsfonts,mathtools, mathrsfs}
\usepackage[toc, page, title]{appendix}
\usepackage{algpseudocode}
\usepackage{algorithmicx}
\usepackage{algorithm}
\algrenewcommand{\algorithmiccomment}[1]{\hskip3em// #1} 

\usepackage{tikz}
\usepackage{amsmath}
\usepackage{soul}

\usepackage{graphicx}
\usepackage{subcaption}
\usepackage{textcomp}
\usepackage{xcolor}

\usepackage[normalem]{ulem}

\usepackage{balance}
\usepackage{graphicx}
\usepackage{enumerate}
\usepackage{etoolbox}
\usepackage{verbatim}
\usepackage{float}
\usepackage{amsthm} 
\usepackage{gensymb}
\usepackage{multirow}

\newbool{IsPrintComment}
  \boolfalse{IsPrintComment}  
\newcommand{\lineCut}[1]
{
    \ifbool{IsPrintComment}
    {
        {\sout{#1}}
    }
    \  
}

\newcommand{\squishlist} 
{
    \begin{list}{$\bullet$}
    {
        \setlength{\itemsep}{0pt}      \setlength{\parsep}{3pt}
        \setlength{\topsep}{3pt}       \setlength{\partopsep}{0pt}
        \setlength{\leftmargin}{1.5em} \setlength{\labelwidth}{1em}
        \setlength{\labelsep}{0.5em}
    }
}

\newcommand{\squishend}
{
    \end{list}
}

\newenvironment{squishedEnumerate}
{\vspace*{-0.2cm}
    \begin{enumerate}
        \setlength{\parskip}{0pt}
        \setlength{\itemsep}{0pt}
        \setlength{\parsep}{3pt}
        \setlength{\topsep}{0pt}
        \setlength{\partopsep}{0pt}
        \setlength{\leftmargin}{1.5em}
        \setlength{\labelwidth}{1em}
        \setlength{\labelsep}{0.5em}

}{\end{enumerate} \vspace*{-0.2cm}}

\newcommand{\ig}[1]
{
    \ifbool{IsPrintComment}
    {
        {\bf \color{cyan} IG 3/13: #1}
    } 
    \  
}
\newcommand{\igc}[1]
{
    \ifbool{IsPrintComment}
    {
        {\color{magenta} #1}
    }
    {
        { #1 }
    }
	\  
}

\newcommand{\shadi}[1]
{
    \ifbool{IsPrintComment}
    {
         {\color{orange}(SAN) #1}
    }
    {
        { #1}
    }
	\  
}

\definecolor{darkgreen}{RGB}{1,150,33}
\newcommand{\shadiS}[1]
{
    \ifbool{IsPrintComment}
    {
         {\color{darkgreen}(SAN suggestion) #1}
    }
    {
        {\color{darkgreen} #1 }
    }
	\  
}

\definecolor{darkyellow}{RGB}{255,220,60}
\newcommand{\shadiC}[1]
{
    \ifbool{IsPrintComment}
    {
         {\color{red} #1}
    }
	\  
}

\newcommand{\todo}[1]
{
    \ifbool{IsPrintComment}
    {
         {\color{red} [#1]}
    }
	\  
}

\definecolor{violet}{RGB}{238,130,238}
\definecolor{purple}{RGB}{180,121,230}
\definecolor{blue-green}{rgb}{0.0, 0.87, 0.87}
\newcommand{\ryc}[1]
{
    \ifbool{IsPrintComment}
    {
        {\color{blue-green}#1}
    }
    {
        {#1}
    }
    \
}

\newcommand{\ry}[1]
{
    \ifbool{IsPrintComment}
    {
         {\color{purple} (RY): #1}
    }
	\ 
}

\newcommand{\needreview}[1]
{
    \ifbool{IsPrintComment}
    {
         {\color{blue} (New Edit): #1}
    }
    {
        { #1 }
    }
    \
}

\newcommand{\review}[1]
{
	\ifbool{IsPrintComment}
	{
		{\color{gray}  #1}
	}
	\
}
\newcommand{\outline}[1]
{
	\ifbool{IsPrintComment}
	{
		{\color{blue}  #1}
	}
	\
}
\newcommand{\sbaTodo}[1]
{
    \ifbool{IsPrintComment}
    {
         {\color{Purple} [SBA: TODO $\rightarrow$ #1]}
    }
	\  
}
\newcommand{\deleteMe}[1]
{
    \ifbool{IsPrintComment}
    {
        
    }
	\  
}
\newcommand{\SBAc}[1]
{
    \ifbool{IsPrintComment}
    {
         {\tt \color{blue} SBA: #1}
    }
	\  
}
\definecolor{darkred}{RGB}{150,50,33}

\newcommand{\SBA}[3]
{
    \ifbool{IsPrintComment}
    {
        {\bf \color{#1} [ SBA : #2 : #3 ]}
    }
    \  
}
\newcommand{\SBAdone}[1]
{
    \ifbool{IsPrintComment}
    {
        {\bf \color{OliveGreen} [ SBA : #1]}
    }
    \  
}
\newcommand{\Print}[2]
{
    \ifbool{IsPrintComment}
    {
        {\color{#1} #2}
    }
	\  
}

\newcommand{\ReplacedWith}[3]
{
    \ifbool{IsPrintComment}
    {
        \st
        {
            #2
        }
        \Print{Orange}
        {
            {\color{#1} #3}
        }
    }
    \  
}
\usepackage{titlesec}

\titlespacing*{\section}
{0pt}{1ex plus 0.2ex minus .2ex}{0.5ex plus .1ex}
\titlespacing*{\subsection}
{0pt}{0.8ex plus 0.2ex minus .1ex}{0.2ex plus .1ex}
\titlespacing*{\subsubsection}
{0pt}{0.7ex plus 0.1ex minus .1ex}{0.2ex plus 0.1ex}

\titleformat*{\subsection}{\fontsize{11}{11}\bfseries}

\newcommand{\name}{SafeHome}

\newcommand{\UIUC}{$^{\dagger}$}
\newcommand{\MS}{$^{\ast}$}

\newcommand{\Figure}{Fig.}
\newcommand{\Figures}{Figs.}
\newcommand{\Section}{Sec.}
\newcommand{\Algorithm}{Algo.}

\newcommand{\Table}{Table}
\newcommand{\Line}{Line}

\newcommand{\Appendix}{Appendix}

\newtheoremstyle{sbaStyle}
  {4pt}
  {4pt}
  {\normalfont}
  {0pt}
  {\bf}
  {:}
  { }
  {}
 
 \theoremstyle{sbaStyle}

\newtheorem{invariant}{Invariant}

\begin{document}
    \date{}
    \title{\Large \bf Home, {\name}: Smart Home Reliability with Visibility and Atomicity }
    
    \author
    {
        {\rm Shegufta Bakht Ahsan{\UIUC}, Rui Yang{\UIUC}, Shadi A. Noghabi{\MS}, Indranil Gupta{\UIUC}}\\
        \{sbahsan2,ry2, indy\}@illinois.edu{\UIUC}, shadi@microsoft.com{\MS}\\
        {\UIUC}University of Illinois at Urbana Champaign. {\MS}Microsoft Research.
    } 


    \maketitle
\vspace*{-1cm}
    \begin{abstract}
        
        Smart environments (homes, factories, hospitals, buildings) contain an increasing number of IoT devices, making them complex to manage. Today, in smart homes where users or triggers initiate  routines (i.e., a sequence of commands), concurrent routines and device failures can cause incongruent outcomes. We describe {\name}, a system that provides notions of atomicity and serial equivalence for  smart  homes. Due to the human-facing nature of smart homes, {\name} offers a spectrum of {\it visibility models} which trade off between responsiveness vs. incongruence of the smart home state. We implemented {\name} and performed workload-driven experiments. We find that a weak visibility model, called {\it eventual visibility}, is almost as fast as today's status quo (up to 23\% slower) and yet guarantees serially-equivalent end states.
        

    \end{abstract}


\section{Introduction}
    \label{sec:safeHome:intro}

    The disruptive smart home market is projected to grow from \$27B to \$150B by 2024 \cite{SmartHomeMarketAnalysis1, SmartHomeMarketAnalysis2}. There is a wide diversity of devices---roughly 1,500 IoT vendors today~\cite{2018MappingMarket}, with the average home expected to contain over 50 smart devices by 2023 \cite{2023Average}. Smart devices cover all aspects of the home, from safety (fire alarms, sensors, cameras), to doors+windows (e.g., automated shades), home+kitchen gadgets, HVAC+thermostats, lighting,  garden sprinkler systems, home security, and others. 
    As the  devices in the home increase in number and complexity, the chances of interactions leading to undesirable outcomes become greater. This diversity and scale is even vaster in other smart environments such as smart buildings, smart factories (e.g., Industry 4.0~\cite{SmartFactory1}), and smart hospitals~\cite{SmartHospital1}.

    Past computing eras---1970s' mainframes, 1990s' clusters, and 2000s' clouds---were successful because of good management systems~\cite{Campbell-Kelly}. What is desperately needed are systems that allow  a group of users to manage their smart home as a single entity rather than a collection of individual devices~\cite{PrinciplesOfSmHmCtrl}. Today, most users (whether in a smart home or a smart factory) control a device using {\it commands}, e.g., turn {\tt ON} a light. Further, major smart home controllers have started to provide users the ability to create {\it routines}. A routine is a  sequence of commands~\cite{routine1, routine2, Transactuations, iRobot:Imprint_whatIs}. Routines are  useful for both: a) convenience, e.g., turn {\tt ON} a group of Living Room lights, then switch on the entertainment system, and b) correct operation, e.g., {\tt CLOSE} window, then turn {\tt ON} AC.

    \noindent {\bf Motivating Examples: } Today's {\it best-effort} way of executing routines can lead to incongruent states in the smart home, and has been documented as the cause of many smart home incidents~\cite{iot-incident, iot-incident2, iot-incident3, E19, Transactuations}~\footnote
    {
        While security issues also abound, we believe such  correctness violations are very common and under-reported as a pain point.
    }. 
    First, consider a routine involving the AC and a smart window~\cite{Fenestra, Velux}:  $R_{cooling}$ { = \{{\tt CLOSE} window; switch {\tt ON} AC\}}. During the execution of this routine, if either the window or the AC fails, the end-state of the smart home will not be what the user desired---either leaving  the window open and AC on (wasting energy), or the window closed and AC off (overheating the home). Another example is a shipping warehouse wherein a robot's routine needs to retrieve an item, package it, and attach an  address label---all these actions are essential to ship the item correctly. In all these cases, lack of {\it atomicity} in the routine's execution  violates the expected outcome.

\begin{figure}[]
        \centering
        \includegraphics[width=0.95\linewidth]{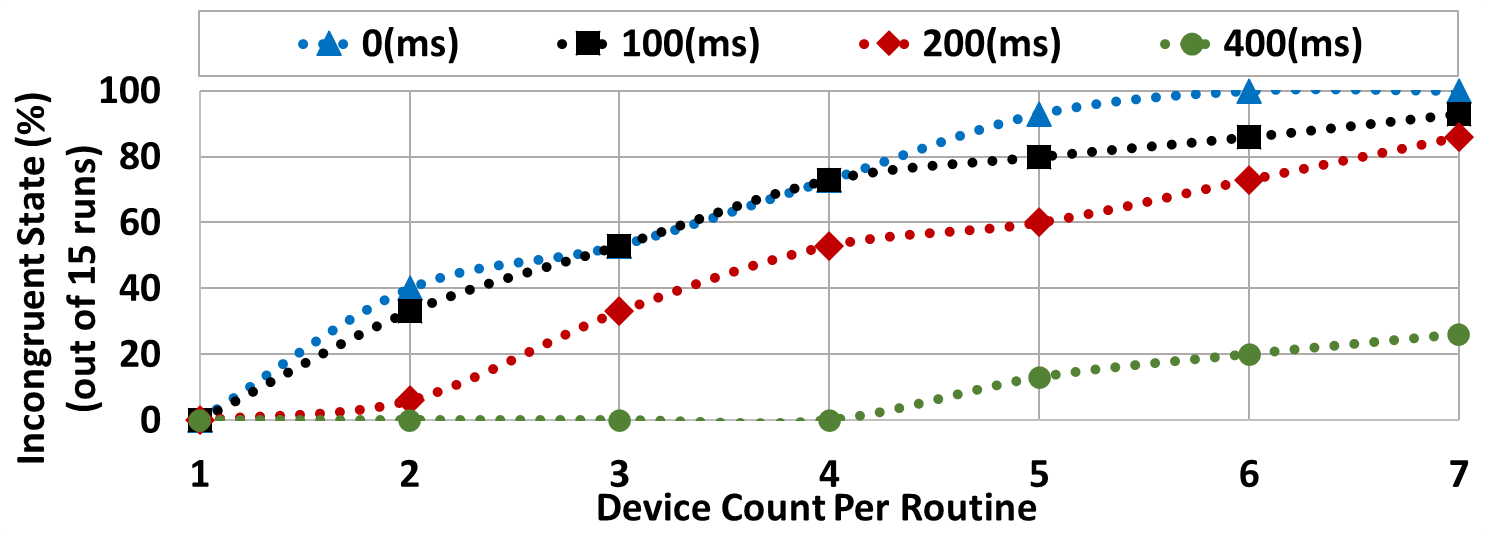}
        \vspace{-0.5em}
        \caption
        {
            \small \it 
            {\bf Concurrency causes Incongruent End-state in a real smart home deployment.}
            Two routines R1 (turn {\tt ON} all lights) and R2 (turn {\tt OFF} all lights) executed on a varying number of devices (x axis),  with routine R2 starting a little after R1 (different lines). Y axis shows fraction of end states that are not serialized (i.e., all {\tt OFF}, or all {\tt ON}). Experiments with TP-Link smart devices~\cite{TP-DevicesInExperiment}.
        }
        \vspace{-1.5em}
        \label{fig:pilotExp}
    \end{figure}

    Our next example deals with concurrent routines. 
    Consider a timed routine $R_{trash}$ that executes every Monday night at 11 pm and takes several minutes to run: $R_{trash}$=\{{\tt OPEN} garage; {\tt MOVE} trash can out to driveway (a robotic trash can like SmartCan \cite{SmartCan}); {\tt CLOSE} garage\}.
    %
    %
   One day the user goes to bed around 11 pm, when she initiates a routine: $R_{goodnight}$=\{switch {\tt OFF} all outside lights; {\tt LOCK} outside doors; {\tt CLOSE} garage\}. Today's state of the art has no {\it isolation} between the two routines, which could result in $R_{goodnight}$ shutting the garage (its last command) while $R_{trash}$ is either executing its first command (open garage), or its second command (moving trash can outside). In both cases,  $R_{trash}$'s execution is incorrect, and equipment may be damaged (garage or trash can). 
 Concurrency even among short routines could result in such incongruences---Figure~\ref{fig:pilotExp} shows such an experiment. The plot shows that two routines simultaneously touching only a few devices cause incongruent outcomes if they start close to each other. In all these cases, {\it isolation semantics} among concurrent routines were not being specified cleanly or enforced. 

    \noindent {\bf Challenges: } This discussion points to the need for a smart home to autonomically provide two critical properties: i) {\it Atomicity} and ii) {\it Isolation/Serializability}. Atomicity ensures that all the commands in a routine have an effect on the environment, or none of its commands do (e.g., if the window is not closed, the AC should not be turned on). Serializability says that the {\it effect} of a concurrent set of routines is equivalent to executing them one by one, in some sequential order, e.g., when  $R_{trash}$ and $R_{goodnight}$ complete successfully,  doors are locked, garage is closed, lights are off, trash can is in the driveway, and no equipment is damaged. 
    
    Specifying and satisfying these two properties in smart homes needs us to tackle certain unique challenges. The first challenge comes from the human-facing nature of the environment. Every action of a routine may be {\it immediately visible to one or more human users}---we use the word ``visible'' to capture any action that could be sensed by any human user anywhere in the smart home.
    This requires us to clearly specify and reason about \underline{visibility models for concurrent routines}. Visibility models provide notions of serial equivalence (i.e., serializability) of routines in a smart home. 
    
    %
    Second, a smart home needs to optimize {\it user-facing metrics}---latency to start the routine, and also latency  to execute it. This motivates us to explore a new \underline{spectrum} of visibility models which trade off the amount of incongruence the user sees {\it during} execution vs. the user-perceived latency, all while guaranteeing  serial-equivalence of the overall execution. Our visibility models are a counterpart to the rich legacy of weak consistency models that have been explored in  mobile systems like  Coda~\cite{coda}, databases  like  Bayou~\cite{bayou} and NoSQL~\cite{eventuallyConsistent}, and shared memory multiprocessors~\cite{adve96}.
        
    Third, in a smart home, {\it device crashes and restarts are the norms}---any device can fail at any time, and possibly  recover later. These failure/recovery events may occur during a command, before a command starts, or after a command has completed. Thus, \underline{reasoning about device failure/restart events} while ensuring atomicity+visibility models is a new challenge. Today's failure handling is either silent or places the burden of resolution on the user. 
        
    Fourth, {\it long-running (or just long) routines are common} in smart homes. A long routine is one that contains at least one {\it long command}.
    A long command exclusively needs to control a device for an extended period, without interruption. Examples include a command to preheat an oven to $400^{\circ}F$, or to run north garden sprinklers for 15 minutes. {Long commands cannot be treated merely  as two short commands, as this would still allow the device to be interrupted by a concurrent routine in the interim, violating isolation. } Long commands need to be treated as first-class commands.

    \noindent {\bf Prior Work:} These challenges have been addressed only piecemeal in literature. Some systems~\cite{HomeOS, DependencyManagement} use priority-based approaches to address concurrent device access. Others~\cite{Rivulet} propose mechanisms to handle failures. A few systems~\cite{ConflictDetection1, DepSys, SIFT} formally verify procedures. 
    {
    Transactuation~\cite{Transactuations} and APEX~\cite{APEX} discuss atomicity and isolation, but their concrete techniques deal with routine dependencies and do not consider users' experience---nevertheless, their mechanisms can be used orthogonally with {\name}. None of the above address atomicity, failures, and visibility together.
    }

    The reader may also notice parallels between our work and the ACID properties  (Atomicity, Consistency, Isolation, and Durability) provided by transactional databases~\cite{DBmngmntSystm}. While other systems like TinyDB~\cite{TinyDB} have drawn parallels between networks of sensors and databases (DBs), the {techniques} for providing ACID in databases do not translate easily to smart homes. The primary reasons are: i) our need to optimize latency (DBs optimize throughput); ii)  device failure (DB objects are replicated, but devices are not, by default); and iii) the presence of long-running routines.

    \noindent {\bf Contributions:}  We present {\it \name}, a management system that provides atomicity and isolation among concurrent routines in a smart environment. For concreteness, we focus the design of {\name} on smart homes  (however, our evaluations look at broader scenarios). {\name} is intended to run at an edge device in the smart home, e.g., a home hub or an enhanced access point. {\name} does not require additional logic on devices; instead, it works directly with the APIs which devices naturally provide (commands are API calls). {\name} can thus work in a smart home containing devices from multiple vendors.

    The primary contributions of this paper are:

    \begin{squishedEnumerate}
        \item A new spectrum of {\it Visibility Models} trading off responsiveness vs. temporary congruence of smart home state.
     
        \item Design and implementation of the {\name} system. 
    
        \item A new way to reason about failures by {\it serializing failure events and restart events into} the serially-equivalent order of routines.
    
        \item New {\it lock leasing} techniques to increase concurrency among routines, while guaranteeing isolation. 
    
        \item Workload-driven experiments to evaluate new visibility models and characterize tradeoffs.
    \end{squishedEnumerate}

     {\name} is best seen as the first step towards a grand challenge. A true OS for smart homes requires tackling myriad problems well beyond what {\name} currently does. These include support for~\cite{hotEdge19}: users to inject signals/interrupts/exceptions, safety property specification and satisfaction, leveraging programming language and verification techniques,
     and in general full ACID-like properties. {\name} is an important building block over which (we believe) these other important problems can then be addressed.
    

\section{Visibility and Atomicity}
    \label{sec:ser_models}

    We first define {\name}'s two key properties--Visibility and Atomicity--and then expand on each.

    \squishlist
        \item {\bf \name-Visibility/Serializability:}
            For simplicity, in this initial part of the discussion we  ignore failures, i.e., we assume devices are always up and responsive. {\name}-Visibility/Serializability means the {\it effect} of the concurrent execution of a set of routines, is identical to an equivalent world where the same  routines all executed serially, in some order. The interpretation of {\it effect} determines different flavors of visibility, e.g., identicality at every point of time, or in the end-state (after all routines complete), or  at critical points in the execution. These choices determine the {\it spectrum} of visibility/serializability {\it models} that we will discuss soon.
      	    
      	\item {\bf \name-Atomicity:}
            After a routine has started, either all its commands have the desired effect on the smart home (i.e., routine {\it completes}), or the system {\it aborts} the routine, resulting in a rollback of  its commands, and gives the user  feedback.

    \squishend  

    \subsection{New Visibility Models in {\name}}
        \label{sec:visbilitymodels}
    
        {\name} presents to the user family a choice in how the effects of concurrent routines are visible. We use the term ``visibility'' to capture all  senses via which a human user, anywhere in the environment, may experience immediate activity of a device, i.e., sight, sound, smell, touch, and taste. Visibility models that are more strict run routines sequentially, and thus may suffer from longer end-to-end latencies between initiating a routine and its completion (henceforth we refer to this simply as {\it latency}). Models  with weaker visibility offer shorter latencies, but need careful  design to ensure the end state of the smart home  is congruent (correct).

        Today's default approach is to execute routines' commands as they arrive, as quickly as possible, without paying attention to serialization or visibility. We call this {\it status quo} model as the {\it Weak Visibility (WV)} model, and its incongruent end states worsen  quickly with scale and concurrency (see \Figure~\ref{fig:pilotExp}). We introduce three  new visibility models.

    \noindent {1. \bf Global Strict Visibility (GSV):}
        In this strong visibility model, {\it the smart home executes at most one routine at any time.} In our {\name}-Visibility definition (\Section~\ref{sec:ser_models}), the {\it effect} for GSV is ``at every point of time'', i.e., every individual action on every device. Consider a 2-family home where one user starts a routine $R_{dishwash}$ {\tt =\{dishwasher:ON; /*run dishwasher for 40 mins*/ dishwasher:OFF;\}}, and another user simultaneously starts a second routine $R_{dryer}${\tt = \{dryer:ON; /*run dryer for 20 mins*/ dryer:OFF;\}}. If the home has low amperage, switching on both dishwasher and dryer  simultaneously may cause an outage (even though these 2 routines touch disjoint devices). If the home chooses GSV, then the execution of $R_{dishwash}$ and $R_{dryer}$ are serialized, allowing at most one to execute at any point of time. Because routines need to wait until the smart home is ``free'', GSV   results in very long latencies to start  routines. In GSV, a  long-running routine also starves other routines.

    \noindent {2. \bf Partitioned Strict Visibility (PSV):}
        PSV is a weakened version of GSV that allows concurrent execution of non-conflicting routines, but limits conflicting routines to execute serially. For instance, for our earlier (GSV) example of  $R_{dishwash}$ and $R_{dryer}$ started simultaneously, if the home has no amperage restrictions, the users should choose PSV--this allows the two routines to run concurrently, and the end state of the home is (serially-) equivalent to the end state if the routines were instead to have been run sequentially
        (i.e., dishes washed, clothes dried). However, if the two routines {\it were} to touch conflicting devices, PSV would execute them serially.
    
    \noindent {3. \bf Eventual Visibility (EV):}
        This is our most relaxed visibility model which  specifies that only {\it when all the  routines have  finished (completed/aborted), the end state of the smart home} is identical to that obtained if {\it all routines were to have been serially executed in some sequential (total) order}. In the definition of {\name}-Visibility, the {\it effect} for EV is the end-state of the smart home after all the routines are finished.

        EV is intended for the relatively-common scenarios where the desired final outcome (of routines) is more important to the users than the ephemerally-visible intermediate states. Unlike GSV, the EV model allows conflicting routines (touching conflicting devices) to execute concurrently--and thus reduces the latencies of both starting and running  routines.

        Consider  two users in a home simultaneously initiating the same routine $R_{breakfast}$
        {\tt =
    		\{
    		    coffee:ON; /*make coffee for 4 mins*/;
    		    coffee:OFF; 
    		    pancake:ON; /*make pancakes for 5 mins*/;
    		    pancake:OFF; 
    		    \}}. 
    		    
    	Both GSV and PSV would serially execute these routines because of the conflicting devices. EV would be able to pipeline them, overlapping the pancake command of one routine with the coffee command of the other routine. EV only cares that at the end both users have their respective coffees and pancakes.

\begin{figure}[]
        \centering
        \includegraphics[width=0.95\linewidth]{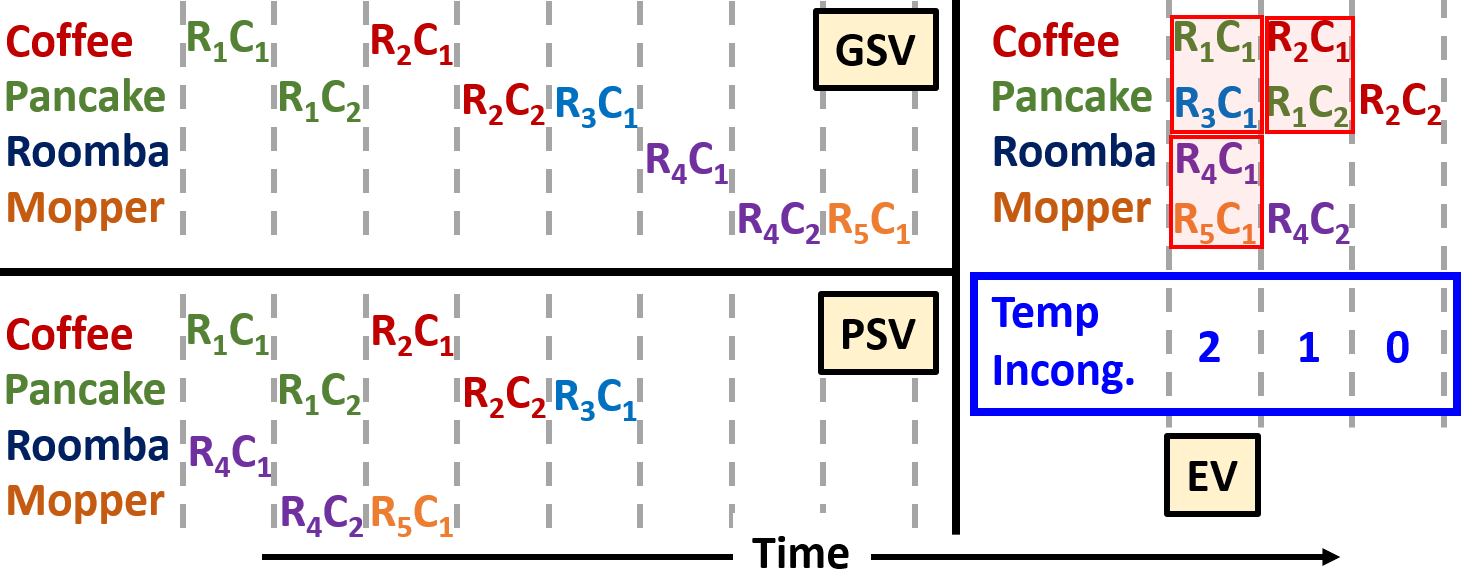}
        \vspace*{-1em}
        \caption
        {
            \small \it 
            {\bf Example routine execution in different visibility models:} 
            a) GSV b) PSV, c) EV. $R_{r}C_{c}$ represents the $c^{th}$ command of the $r^{th}$ routine. 
            In EV, red boxes show a pair of incongruent commands and the blue box shows the total number of temporary incongruences.}
        \vspace{-1.5em}
        \label{fig:realDeploymentLineageTable}
    \end{figure}

    
    \noindent\textbf{Common Example -- 3 Visibility Models:}         
        \Figure~\ref{fig:realDeploymentLineageTable} shows an example with 5 concurrent routines executed for our three visibility models. This is the outcome of a real run of {\name} running on a Raspberry Pi, over 5 devices connected via TP-Link HS-105 smart-plugs~\cite{_TP-HS105}. The routines are:

        {\noindent}{\it $R_1$: makeCoffee(Espresso); makePancake(Vanilla);}\\
        {\it $R_2$: makeCoffee(Americano);  makePancake(Strawberry); }\\
        {\it $R_3$: makePancake(Regular); }\\
        {\it $R_4$: startRoomba(Living room);  startMopping(Living room); }\\
        {\it $R_5$: startMopping(Kitchen); }
        
        GSV takes the longest execution time of 8 time units as it serializes execution. PSV reduces execution time to 5 time units by parallelizing unrelated commands, e.g., $R_1$'s coffee command  and $R_4$'s Roomba command at time $t=0$. EV is the fastest, finishing all routines by 3 time units. Average latencies (wait to start, wait to finish) are also fastest in EV, then PSV, then GSV. 
        The figure shows that EV exhibits ``temporary incongruence''--routines whose intermediate state is not serially equivalent. 
        EV guarantees a temporary incongruence of zero when the last routine finishes.

        \Table~\ref{tab:visibility-models} contrasts the properties of the four visibility models.  \Table~\ref{tab:safehome-examples} summarizes the examples discussed so far.

       \begin{table}[]
    	\resizebox{\linewidth}{!}
    	{
    		\begin{tabular}{|p{0.2\linewidth}|p{0.2\linewidth}|p{0.34\linewidth}|p{0.25\linewidth}|p{0.23\linewidth}|}
    			\hline
    			& {\bf GSV}               	& {\bf PSV }		& {\bf EV }				& {\bf WV}   \\ \hline
    			{\bf \it Concurrency} 	& At most one routine 
    			& Non-conflicting routines  concurrent
    			& Any serializable routines concurrent
    			& Any routines concurrent \\ \hline
    			{\bf \it End State} 	& Serializable
    			& Serializable
    			& Serializable
    			& Arbitrary \\ \hline
    			{\bf \it Wait Time: time to start routine}  & High
    			& High for conflicting routines, low for non-conflicting routines
    			& Low for all routines (modulo conflicts)
    			& Low for all routines \\ \hline
    			%
    			{\bf \it User \newline Visibility} & Congruent at all times 
    			& Congruent at end, and at start/complete points of routines 
    			& Congruent at end 
    			& May be incongruent at anytime or end (Fig.~\ref{fig:pilotExp}) \\ \hline
    			%
    			%
 
    		\end{tabular}
    	}
    	\vspace*{-0.5em}
    	\caption{\bf Spectrum of  Visibility Models in \name.}
    	\vspace*{-1.5em}
    	\label{tab:visibility-models}
    \end{table}
    
\begin{table*}[]
    \centering
	\small
	\resizebox{\linewidth}{!}{
		\begin{tabular}{|p{0.52\linewidth}|p{0.55\linewidth}|p{0.14\linewidth}|p{0.01\linewidth}|}
			\hline
		{\bf Example Routines} & {\bf Scenario and Possible Behavior} & \multicolumn{2}{|p{0.17\linewidth}|}{{\bf \name{} Feature}} \\ \hline
		``cooling''{\tt =\{window:CLOSE; AC:ON;\}}	
		& If executed partially, can leave window open and AC on (wasting energy) or the window closed and AC off (overheating home). 
		& \multicolumn{2}{|p{0.19\linewidth}|}{ Atomicity} \\ \hline
		``make coffee''=
		{\tt \{coffee:ON; /*make coffee for 4 mins*/ ; coffee:OFF;\} }
		& Coffee maker should not be interrupted by another routine. 
		E.g, user-1 invokes make coffee, and in the middle, user-2 independently invokes make coffee.
		& \multicolumn{2}{|p{0.19\linewidth}|}{Long running routines \& mutually exclusive access}\\ \hline

	
	    {\tt $R_1$=\{dishwasher:ON; (dishwasher runs for 40 mins); dishwasher:OFF;\}}\newline
	    $R_2$={\tt \{dryer:ON; (dryer runs for 20 mins); dryer:OFF;\} }
	    &  If home has low amperage, simultaneously running two power-hungry devices may cause outage (GSV).
		& \multicolumn{2}{|p{0.19\linewidth}|}{ Global Strict Visibility (GSV)}\\ \hline
		{\tt $R_1$=\{coffee:ON; /*make coffee for 4 mins*/; coffee:OFF;\} \newline
		$R_2$=\{lights:ON, fan:ON\} }
		&  Two routines touching disjoint devices should not block each other (PSV). 
        & \multicolumn{2}{|p{0.19\linewidth}|}{Partitioned Strict Visibility (PSV), closest to \cite{Transactuations}}\\ \hline
		``breakfast''{\tt=\{coffee:ON; /*make coffee for 4 mins*/;  coffee:OFF, 
		pancake:ON; /*make pancakes for 5 mins*/;  pancake:OFF; 
		\} }
		&
		Two users can invoke this same routine simultaneously. The two routines can be pipelined thus allowing some concurrency without affecting correctness (EV). (Both GSV and PSV would have serialized them.) 
        & \multicolumn{2}{|p{0.19\linewidth}|}{Eventual Visibility (EV)} \\ \hline
		
	    ``leave home''{\tt =\{lights:OFF (Best-Effort);  door:LOCK;\} }
		& Requiring all commands to finish too stringent, so only second command is Must (required). If light unresponsive, door must lock, otherwise routine aborts. 
		& \multicolumn{2}{|p{0.19\linewidth}|}{ Must and Best-Effort commands}\\ \hline
		{\tt ``manufacturing pipeline'' with \textit{k} stages and \{$R_1, R_2, ..., R_k$\} routines}
		& If any stage fails, entire pipeline must stop immediately. 
		& Strong GSV serialization (S-GSV)
		& \multirow{4}{*}[-1ex]{\rotatebox[origin=c]{270}{Failure Serialization}}\\ \cline{1-3} 
		``cooling''{\tt =\{window:CLOSE; AC:ON;\} }
		& If \textit{anytime} during the routine (from start to finish), the AC fails or window fails, the routine is aborted. 
		& Loose GSV serialization (GSV) &\\ \cline{1-3}
	    ``cooling''{\tt =\{window:CLOSE; AC:ON;\} }
		& If window fails after its command {\it and }  remains failed at finish point of routine, routine is aborted. 
		& PSV serialization &\\ \cline{1-3}
        ``cooling''{\tt =\{window:CLOSE; AC:ON;\} }
		& If window fails after it is closed (but before AC is accessed), routine  completes successfully--window failure can be serialized after routine. 
		& EV serialization &\\ \hline
		\end{tabular}
	}
	\caption
    {
       \bf Example scenarios in a smart home, and  {\name}'s corresponding features. 
    }
	\label{tab:safehome-examples}
	\vspace*{-0.5cm}
\end{table*}

    \subsection{\name-Atomicity}
        \label{sec:safeHomeAtomicity}
         
        \name-Atomicity states that after a routine has started, either all its commands have the desired effect on the smart home (i.e., routine {\it completes}), or the system {\it aborts} the routine, resulting in a rollback of  its commands, and gives the user  feedback. 
        Due to the physical effects of smart home routines, we discuss three deviations from traditional notions of atomicity. 

            First, we allow the user to tag some commands as {\it best-effort}, i.e., optional. A routine is allowed to complete successfully even if any best-effort commands fail. Other commands, tagged as {\it must}, are required for routine completion---if any {\it must} command fails, the routine must abort. This tagging acknowledges the fact that users may not consider all commands in a routine to be equally important. For instance, a   ``leave-home-for-work'' routine may contain commands which lock the door (must commands) and  turn off  lights (best-effort commands)---even if the lights are unresponsive, the doors must still lock. The user receives feedback about the failed best-effort  commands, and she is free to either ignore or re-execute them. 

            Second, aborting a routine requires undoing  past-executed commands. Many commands can be rolled back cleanly, e.g., command {\tt turn Light-3  ON} can be undone by \name{} issuing a command setting Light-3 to {\tt OFF}. A small fraction of commands is impossible to physically undo, e.g., {\tt run north sprinklers for 15 mins}, or {\tt blare a test alarm}. For such commands, we undo by restoring the device to its state before the aborted routine (e.g., set the sprinkler/alarm state to {\tt OFF}).  Alternately, a user-specified undo-handler can be used.

            Finally, we note that when a routine aborts, {\name} provides feedback to the user (including logs), and she is free to either re-initiate the routine or ignore the failed routine.

    \section{Failure Handling and Visibility Models}
    \label{sec:failureHandling}
    Smart home devices could  fail or become unresponsive, and then later restart. {\name} needs to reason cleanly about failures or restarts that occur during the execution of concurrent routines. 
    We only consider fail-stop and fail-recovery models of failures in the smart home (Byzantine failures are beyond our scope). 
        
    Because device failure events and restart events are visible to human users, our visibility models need to be amended. Consider a device $D$ which routine $R$ touches via one or more commands. $D$ might fail {\it during} a command from $R$, or {\it after} its last command from $R$, or {\it before} its first command from  $R$, or {\it in between } two commands from $R$. A naive approach may be to abort routine $R$ in all these cases. However, for some relaxed visibility models like Eventual Visibility, if the  failure event occurred anytime after completing the device's last command from $R$, then the event could be serialized to occur {\it after} the routine $R$ in the serially-equivalent order (likewise for a failure/restart before the first command to that device from $R$, which can be serialized to occur before $R$).

    Thus a key realization in {\name} is that we need to  {\it serialize failure events and restart events alongside  routines themselves}. We can now restate the {\name}-Visibility property from \Section~\ref{sec:ser_models}, to account for failures and restarts:
        
    {\bf \noindent {\name}-Visibility/Serializability (with Failures and Restarts):}
            The {\it effect} of the concurrent execution of a set of routines, occurring along with  concurrent device failure events and device restart events, is identical to an equivalent world where the same routines, device failure events, and device restart events, all occur sequentially, in some order~\footnote{This idea has analogues to distributed systems abstractions such as view/virtual synchrony, wherein failures and multicasts are totally ordered~\cite{VSync}. 
            We do not execute multicasts in the smart home.}

    First, we define the failure/restart event to be the event when the edge device (running {\name}) {\it detects} the failure/restart (this may be different from the actual time of failure/restart). 
    {
    Second, failure events and restart events {\it must} appear in the final serialized order. On the contrary, routines may appear in the final serialized order (if they complete), or not appear (if they abort). 
    }
    We next reason explicitly about failure serialization for each of our visibility models from \Section~\ref{sec:visbilitymodels}. 
     
     \vspace{4pt}  
    \noindent {\bf 1. Failure Serialization in Weak Visibility:}
        Today's Weak Visibility has no failure serialization. Routines affected by failures/restarts  complete and cause incongruent end-states.
        
     \vspace{4pt}    
    \noindent {\bf 2. Failure Serialization in Global Strict Visibility:}
        Because GSV intends to present the picture of a single serialized home to the user, { if {\it any} device failure event or restart event were to occur while a  routine is executing (between its start and finish), the routine must be aborted}. There are two sub-flavors herein:  (A) {\it Basic GSV or Loose GSV (GSV)}: Routine aborts only if it contains at least one command that touches failed/restarted device; (B) {\it Strong GSV (S-GSV)}: Routine aborts even if it does not have a command that touches failed/restarted device. A routine $R_{shade}$ on living room shades can complete, if master bathroom shades fail, in GSV but not S-GSV. In S-GSV, the final serialization order contains the failure/restart event but not the aborted routine $R_{shade}$. In GSV, the final serialization order contains both $R_{shade}$ (which completes) and the failure/restart event, in arbitrary order.
    
     \vspace{4pt}    
    \noindent {\bf 3. Failure Serialization in Eventual Visibility:} 
        For a given set of routines (and concurrent failure events and restart events), the {\it eventual (final)}  state of the actual execution is equivalent to the end state of a world wherein the final successful routines, failure device events, and failure restart events, all occurred in some serial order.
    
        Consider routine $R$, and the failure event (and potential restart event) of one device $D$. Four cases arise:
        
        \begin{squishedEnumerate}
        
             \item If $D$ is not touched by $R$, then $D$'s failure event and/or restart event can be arbitrarily ordered w.r.t. $R$.
             
             \item If $D$'s failure and restart events both occur {\it  before  $R$ first touches the device}, then the failure and restart events are {\it serialized  before $R$}.
            
            \item If $D$'s failure event occurs {\it after the last touch of $D$ by $R$}, then $D$'s failure event (and eventual restart event) are {\it  serialized after $R$}.
            
            \item In all other cases, routine $R$ aborts due to $D$'s failure. $R$ does not appear in the final serialized  order.
        \end{squishedEnumerate}
    \noindent These are applicable to each concurrent routine accessing $D$.
    
     \vspace{4pt}    
    \noindent {\bf 4. Failure Serialization in Partitioned Strict Visibility:} 
        This is a modified version of EV where we change condition 3 (from 1-4 in EV above) to the following:
    
    \noindent {\it 3*.} If $D$'s failure event occurs {\it after the last touch of $D$ by $R$}, and {\it has recovered when $R$ reaches its finish point}, then $D$'s failure event and  restart event are {\it  serialized right after $R$}.

{

    \noindent{\bf Example---Effect of Failure on Three Visibility Models:} Consider the routine from Section~\ref{sec:safeHome:intro}, $R_{cooling}$:{\tt = \{{\tt CLOSE} window; switch ON AC;\}}. Suppose the ``window'' device fails concurrently with the routine (between its start and finish times). GSV always aborts $R_{cooling}$  regardless of when the window failed. PSV aborts $R_{cooling}$ only if the window remains failed at $R_{cooling}$'s finish point. EV does {\it not} need to abort $R_{cooling}$ if window fails any time after $R_{cooling}$'s first command has completed successfully, even if window remains failed at $R_{cooling}$'s finish time. 
    EV places the window failure event after $R_{cooling}$ in the serialization order, and the smart home's end state is equivalent. If the window fails and restarts before $R_{cooling}$'s first command, EV serializes the failure and restart before $R_{cooling}$, and executes $R_{cooling}$ correctly.   Thus EV has the least chance of aborting a routine due to a failure.

    Table \ref{tab:safehome-examples} summarizes all our examples so far and
    \Figure~\ref{fig:serializingFailure} summarizes our failure handling rules.
    

\begin{figure}[]
    \centering
    \includegraphics[width=0.9\columnwidth]{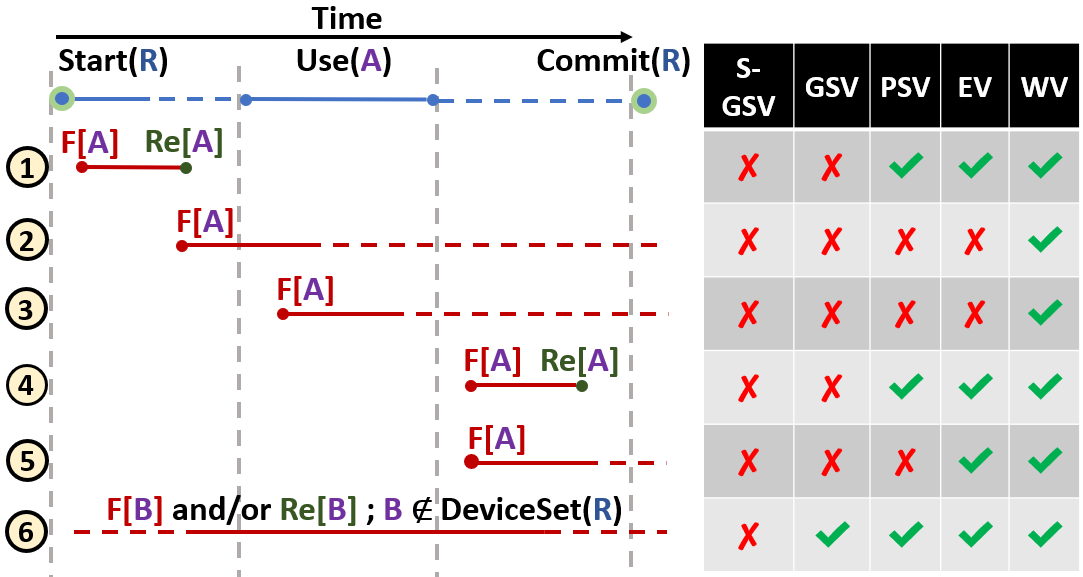}
    \vspace*{-0.5em}
    \caption
    {\small \it 
        {\bf Failure Serialization: 6 cases, and their handling in  Visibility Models.}
        {\textcolor{teal} \checkmark} - execute routine,
        {\textcolor{red} X } - abort routine.
        At {\it F[A] }/{\it Re[A]} the edge device detects the failure/restart (resp.) of device A.}
     \vspace{-1.5em}
    \label{fig:serializingFailure}
\end{figure}

}

\section{Eventual Visibility: {\name} Design}
    \label{sec:safeHome_Design}
     
    In order to maintain correctness for Eventual Visibility (i.e., serial-equivalence), {\name} requires routines to lock devices before accessing them. Because long routines can hold locks and block short routines, we introduce {\it lock leasing} across routines (\Section~\ref{sec:locks}). This information is stored in the {\it Locking Data-structure} (\Section~\ref{sec:locksDS}). The {\it lineage table} ensures invariants required to guarantee Eventual Visibility (\Section~\ref{sec:lineage}).

    \subsection{Locks and Leasing}
        \label{sec:locks}
    
        \noindent{\bf {\name} prefers Pessimistic Concurrency Control (PCC):}
            {\name} adopts pessimistic concurrency control among routines, via (virtual) locking of devices. Abort and undo of routines are disruptive to the human experience, causing (at routine commit point) rollbacks of device states across the smart home. Our goal is to minimize abort/undo only to situations with device failures, and avoid aborts  because routines touch conflicting devices. Hence we eschew optimistic concurrency control approaches and use locking~\footnote{For the limited scenarios where routines are known to be conflict-free, optimistic approaches may be worth exploring in future work.}. 
            
            {\name} uses {\it virtual locking} wherein each device has a virtual lock (maintained at the edge device running \name), which must be acquired by a routine before it can execute any command on that device. A routine's lock acquisition and release do not require device access, and are not blocked by device failure/restart. 
        
            In order to prevent a routine from aborting midway because it is unable to acquire a lock, {\name} uses {\it early lock acquisition}---a routine acquires, at its start point, the locks of all the devices it wishes to touch. If any of these acquisitions fails, the routine  releases all its locks immediately and retries lock acquisition. Otherwise, acquired locks are released (by default) only when the routine finishes.
            
            \noindent{\bf Leasing of Locks: } To minimize chances of a routine being unable to start because of locks held by other routines, \name{} allows routines to lease locks to each other. 
            Two cases arise: 1)  routine $R_1$ holds the lock of device $D$ for an extended period {\it before $R_1$'s first access} of $D$, and 2)  $R_1$ holds the lock of device $D$ for an extended period {\it after $R_1$'s last access} of $D$. Both cases prevent a concurrent routine $R_2$, which also wishes to  access $D$, from starting.
            
            {\name} allows a routine $R_{src}(=R_1)$ holding a lock (on device $D$) to {\it lease the lock} to another routine $R_{dst}(=R_2)$. When $R_{dst}$ is done with its last command on $D$, the lock is returned back to $R_{src}$, which can then normally use it and release it. 
            We support two types of lock leasing:
      
            \squishlist
                \item {\bf Pre-Lease:}
                    $R_{src}$ has started but has not yet accessed $D$. A lease at this point to $R_{dst}$ is called a {\it pre-lease}, and places $R_{dst}$ {\it ahead} of $R_{src}$ in the  serialization order. After $R_{dst}$'s last access of $D$, it returns the lock to $R_{src}$. If $R_{src}$ reaches its first access of $D$ before the lock is returned to it, $R_{src}$ waits. After the lease ends, $R_{src}$ can use the lock normally.
    
                \item {\bf Post-Lease:}
                    $R_{src}$ is done accessing device $D$, but the routine itself has not finished yet. A lease at this point to $R_{dst}$ is called a {\it post-lease}, and places $R_{dst}$ {\it after} $R_{src}$ in the serialization order. If $R_{src}$ finishes  before $R_{dst}$, the lock ownership is permanently transferred to $R_{dst}$. Otherwise, $R_{dst}$ returns the lock when it finishes.
    		\squishend
    
            A prospective pre/post-lease is disallowed if a previous action (e.g., another lease) has already determined a serialization order between $R_{src}$ and $R_{dst}$ that would be contradicted by this  prospective lease. In such cases $R_{dst}$ needs to wait until $R_{src}$'s normal lock release. 
            Further, a post-lease is  not allowed if at  least one device $D$  is written by $R_{src}$ and then  read by $R_{dst}$. This prevents \name{} from suffering dirty reads from aborted routines. We prevent scenarios like this--$R_{src}$ switches on a light, and $R_{dst}$ has a conditional clause based on that light's status, but $R_{src}$ subsequently aborts. Cascading aborts are handled in \cite{Transactuations}, whose techniques can be used orthogonally with ours.
    
            To prevent starvation, i.e., from $R_{src}$  waiting indefinitely for the returned lock, leased locks are revoked after a timeout. The timeout is calculated based on the  estimated time between $R_{dst}$'s first and last actions on $D$, multiplied by a leniency factor (we use $1.1\times$). Lock revocation before $R_{dst}$'s last access of $D$ causes $R_{dst}$ to abort.

    \subsection{Locking Data-structure}
        \label{sec:locksDS}

        \begin{figure}[]
        	\centering
        	\includegraphics[width=0.8\columnwidth]{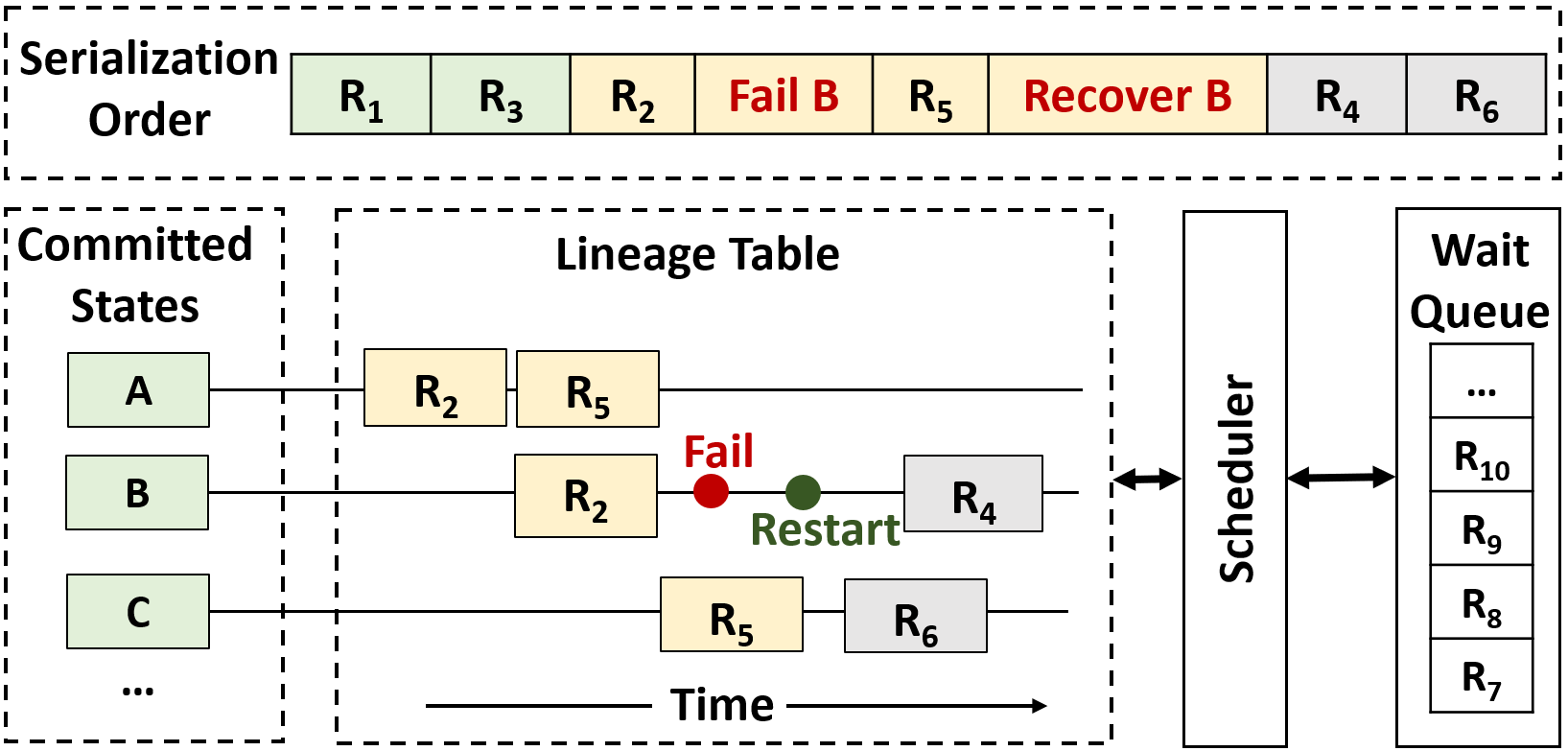}
        	\caption
        	{
        		\small \it
        		{\bf {\name}'s Architecture for Eventual Visibility.}
        	}
        \vspace{-1.2em}
        \label{fig:overall-datastructures}
    	\end{figure}
    	
    {\name} adopts a state machine approach~\cite{SMR90} to track  current device states, future planned actions by routines, and a serialization order. 
		{\name} maintains, at the edge device (e.g., Home Hub or smart access point), a {\it virtual locking table data-structure} (\Figure~\ref{fig:overall-datastructures}). This contains:
    	
    	\squishlist
            \item {\it Wait Queue:} Queue of  routines initiated but not started. When a routine is added, it is assigned an incremented  routine ID.
            
            \item {\it Serialization Order:} Maintains the current serialization order of routines, failure events, and restart events. For completed routines (shaded green), the order is finalized. All other orders are tentative and may change, e.g., based on lock leases. Failure and restart events may be moved flexibly  among unfinished routines.

            \item {\it Lineage Table:} Detailed in Section~\ref{sec:lineage}, this maintains, for each device, a {\it lineage}: the {\it planned} transition order of that device's lock.
            
            \item {\it Scheduler:} Decides when routines from Wait Queue are started, acquires locks, and maintains serialization order.
            
            \item {\it Committed States:} For each device, keeps its last committed state, i.e., the effect of the last successfully routine. This may be different from device's actual state, and is needed to ensure serialization and rollbacks under aborts.
              
        \squishend

    \subsection{\bf Lineage Table}
        \label{sec:lineage}
        \begin{figure}[]
        	\centering
        	\includegraphics[width=0.9\columnwidth]{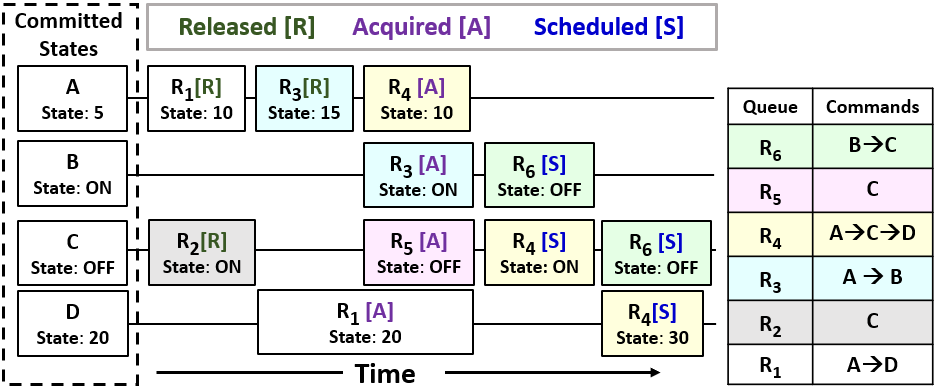}
        	\caption
        	{\small
        		{
        		    \bf
        		    Sample Lineage Table, with 6 routines. Some fields are omitted for simplicity.
        		}
        	}
        	\vspace{-1.2em}
        	\label{fig:lockPattern}
        \end{figure} 
        
        The {\it lineage} of a device represents a temporal plan of when the device will be acquired by concerned routines. The lineage of a device starts with its latest committed state, followed by a sequence of {\it lock-access} entries (\Figure~\ref{fig:lockPattern})--these are ``stretched'' horizontally. A width of a lock-access entry represents how long that routine will acquire that lock. Each lock-access entry for device $D$ consists of:
        {\it i.} A routine ID,
        {\it ii.} Lock {\it status} ({\tt Released, Acquired, Scheduled})
        {\it iii.} Desired device state by the command (e.g., {\tt ON/OFF})
        and {\it iv.} Times: a start time ($T_{start}(R_i)$), and duration ($\tau_{R_i}(D)$) of the lock-access. 

		In the example of \Figure~\ref{fig:lockPattern}, a {\tt Scheduled [S]} status indicates that the routine is scheduled to access the lock. An {\tt Acquired [A]} status shows  it is holding and using the lock. A {\tt Released [R]} status means  the routine has released the lock. 
	
		The duration field, $\tau_{R_i}(D)$, is set either based on known time to run a long command (e.g., run sprinkler for 15 mins), or an estimate of the command execution time. Our implementation uses a fixed $\tau_{R_i}(D)=\tau_{timeout}$ for all short commands (100ms based on our experience). $\tau_{R_i}(D)$ is also used to determine the revocation timeout for leased locks, along with a multiplicative leniency factor (1.1 in our implementation).
   		
      	To maintain serializability, four key invariants are assured:
      	
        \begin{invariant}[{\bf Future Mutual Exclusion: 
        Lock-accesses in a device's lineage list do not overlap in time}]
            \label{inv:non_overlapping_lock_access}
            No device is planned to be locked by multiple routines. Gaps in its lineage list indicate times the device is free.
        \end{invariant}
        
        \begin{invariant}[{\bf Present Mutual Exclusion: At most one {\tt Acquired} lock-access exists in each lineage list}]
            \label{inv:atmost_one_Acquired}
            No device is locked currently by multiple routines.
        \end{invariant}
        
        \begin{invariant}[{\bf Lock-access [R]$\longrightarrow$[A]$\longrightarrow$[S]}]
            \label{inv:lock_access}
            In each  lineage list, all {\tt Released} lock-access entries occur to the left of (i.e., before) any {\tt Acquired} entries, which in turn appear to the left of any {\tt Scheduled} entries.
        \end{invariant}
        
        \begin{invariant}[{\bf Consistent ``serialize-before'' ordering among lineages}]
            \label{inv:serialize_before}
            Given two routines $R_i, R_j$, if there is at least one device $D$ such that: lock-access$_{D} (R_i)$ occurs to the left of lock-access$_{D} (R_j)$ in $D$'s lineage list, then for every other device $D'$ touched by both $R_i, R_j$, it is true that: lock-access$_{D'} (R_i)$ occurs to the left of  lock-access$_{D'} (R_j)$. Hence $R_i$ is  \textit{serialized-before} $R_j$.
        \end{invariant}


   	    \noindent{\bf Transition of Lock-accesses:}
            The status of lock-accesses  changes upon  certain events. First, when a routine's last access to a device ends, the {\tt Acquired} lock-access ends, and transitions to {\tt Released}. The next {\tt Scheduled} lock-access turns to {\tt Acquired}: i) either immediately (if no gap exists, e.g., $R_4$ after $R_5$ releases $C$ in \Figure~\ref{fig:lockPattern}), or ii) after the gap has passed, e.g., $R_4$ after $R_1$ releases $D$ in \Figure~\ref{fig:lockPattern}.

            Second, when scheduling a new routine $R$ (from the wait queue), a {\tt Scheduled} lock-access entry is added to all device lineages that $R$ needs (e.g.,  $R_6$ in \Figure~\ref{fig:lockPattern} adds lock-accesses for  B and C).
            Third, when a routine finishes (completes/aborts), all its lock-access entries are removed, releasing said locks. If the routine completed successfully,  committed states are updated. For an abort,  device states are rolled back.

	
        \begin{figure}[]
        	\centering
        	\includegraphics[width=0.85\linewidth]{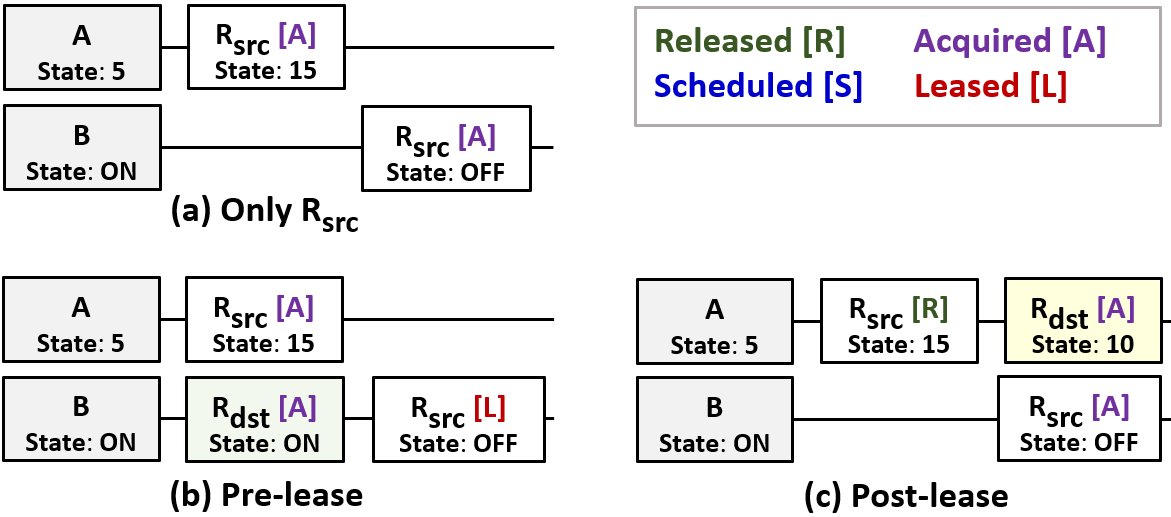}
        	\caption
        	{
        		\small \it 
        		{\bf Lineage table with  Lock Leasing}. a) Lineage before leasing with only $R_{src}$, b) Pre-lease to $R_{dst}$ that only accesses device B, and c) Post-lease to $R_{dst}$ that only accesses device A.
        	}
        	\vspace{-1.5em}
        	\label{fig:prePostLease}
        \end{figure}

	    \noindent{\bf Leasing of Locks:}
	        Consider a pre-lease from $R_{src}$ to $R_{dst}$ (\Figure~\ref{fig:prePostLease}(b)). First, a new {\tt Acquired} lock-access for $R_{dst}$ is placed \textit{before} (to the left of) the lock-access of $R_{src}$ in the lineage table. Second, the lock-access of $R_{src}$ is changed to ``Leased ($R_{dst}$)'' status. 
            
            Figure \ref{fig:prePostLease}(c) shows a post-lease: a new {\tt Acquired} lock-access of $R_{dst}$ is placed \textit{after} (to the right of) the lock-access of $R_{src}$ and the lock-access of $R_{src}$ changes to {\tt Released}. 
	    
        \noindent{\bf Aborts and Rollbacks:}
            For an aborted routine $R_i$, we roll back states of only those devices $D$ in whose lineage $R_i$ appeared. For a device $D$, there are two cases:

            \squishlist
                \item {\it Device $D$ was last {\tt Acquired} by routine $R_j$ ($\neq R_i$):}
                We remove $R_i$'s lock-access from $D$'s lineage. This captures two possibilities: a) $R_i$ never executed actions on $D$ (e.g., \Figure~\ref{fig:lockPattern}: device C when aborting $R_4$), or b) $R_i$ leased $D$ to another routine $R_j$, and since $R_i$ is aborting, $R_j$'s effect will be the latest  (e.g., \Figure~\ref{fig:lockPattern}: device A when aborting $R_1$). 
                
                \item {\it Device $D$ was last {\tt Acquired} by routine $R_i$}
                (e.g. device C when aborting $R_5$ in \Figure~\ref{fig:lockPattern}): We: 1) remove the $R_i$'s lock-access  from $D$'s lineage, and  2)  issue a command to set $D$'s  status to $R_i$'s  {\it immediately left/previous}  lock-access entry in the lineage  (if none exist, use Committed State), unless the device is already in this desired state. 
            \squishend
    
\begin{figure}[]
        	\centering
        	\begin{subfigure}[b]{0.49\columnwidth}
        		\includegraphics[width = \columnwidth]{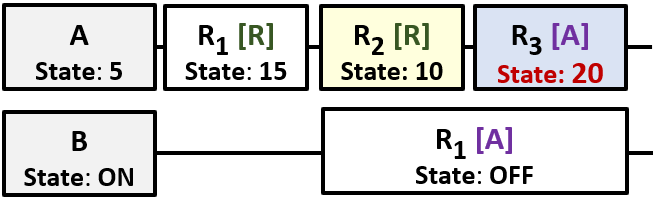}
        		\caption{Before commit}
        		\label{fig:commit_before}
        	\end{subfigure}
        	\begin{subfigure}[b]{0.49\columnwidth}
        		\includegraphics[width = \columnwidth]{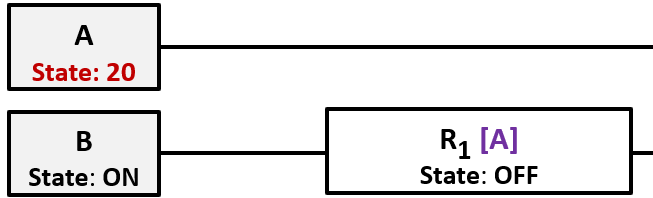}
        		\caption{After $R_3$ commits}
        		\label{fig:commit_after}
        	\end{subfigure}
        	\caption
        	{
                \small \it 
                {\bf Commit with compaction.}
            }
            \label{fig:commit}
        \end{figure}

    	\noindent{\bf Committing (Successfully Completing) a routine:}
            When a routine reaches its finish point, it commits (completes successfully) by: i) updating Committed States, and ii) removing its lock-access entries. $R_j$ might appear after $R_i$ in the serialization order but complete earlier, e.g.,  due to lock leasing. {\name} allows such routines to commit right away by using  {\it commit compaction}--routines later in the serialization order will overwrite effects of earlier routines (on conflicting devices). This is similar to ``last writer wins'' in NoSQL DBs~\cite{eventuallyConsistent}. Concretely, for all common devices we remove both $R_i$'s lock-access, and all lock-accesses before it  (\Figure~\ref{fig:commit}). 
            
\begin{figure}[]
    			\centering    			\includegraphics[width=0.8\columnwidth]{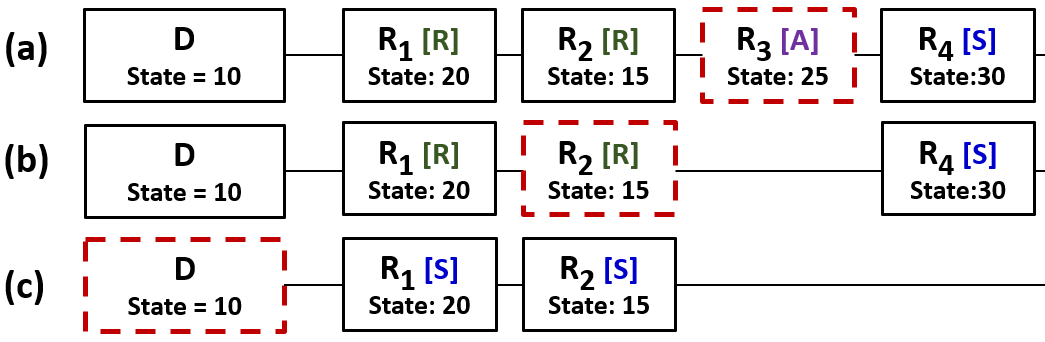}
    			\caption
    			{
    				\small \it 
    				{\bf Inferring the current device status}.
    				The dashed boxes point to the current device status in three different scenarios.
    			}
    		\vspace{-1.2em}
    		\label{fig:design:currentDevStatus}
    		\end{figure}

        \noindent{\bf Current Device Status:} 
            A device's current status is needed at several points, e.g., abort. Due to uncompleted routines, the actual status may  differ from the committed state. 
            The lineage table suffices to estimate a device's current state (without querying the device). \Figure~\ref{fig:design:currentDevStatus} shows the three different cases. (a) If an {\tt Acquired} lock-access entry   exists, use it (e.g., $R_3$ in \Figure~\ref{fig:design:currentDevStatus}(a) with $D=25$ ).
            (b) Otherwise, if lock-accesses exist with lock status {\tt Released}, use the right-most entry  (e.g., $R_2$ in \Figure~\ref{fig:design:currentDevStatus}(b) with $D=15$). (c) Otherwise, use the Committed State entry  (e.g., committed state $D=10$ in \Figure~\ref{fig:design:currentDevStatus}(c)). 
            
    \section{Scheduling Policies for Eventual Visibility}
\label{sec:EVdesign}

         When a new routine arrives, {\name} needs to  ``place'' it in the serialization order, adhering to invariants of \Section~\ref{sec:lineage}. This is the scheduling problem. We present three alternatives.

        \noindent\textbf{First Come First Serve (FCFS) Scheduling:}
        Routines are serialized in  order of arrival. When a routine arrives, its lock-access entries are {\it appended} to the lineage table. FCFS avoids pre-leases as they would violate serialization order. Post-leases are allowed. 
    
        FCFS is attractive if a user expects routines to execute in the order they were initiated. However, FCFS  prolongs time between routine submission and start. 
        
            

        \noindent\textbf{Just-in-Time (JiT) scheduling:}
         JiT greedily places a new routine at the {\it earliest position (in the lineage) when it is eligible to start.} 
        %
	    %
	    {
	    JiT triggers an \textit{eligibility test} upon either: (i) each routine arrival, or (ii) on every lock release. 
	    The eligibility test greedily checks for  routine $R$ if it can now  acquire all its locks, either right away, or via pre-leases or post-leases. For case (ii) we run the eligibility test only on those waiting routines that desire the released device. To mitigate starvation, we use a per-routine TTL (Time To Live)---when a waiting routine $R$'s TTL expires, $R$ is prioritized to start next (ties broken by arrival order). 
	   }

\begin{figure}[]
        	\centering
        	\begin{subfigure}[b]{0.32\columnwidth}
        		\includegraphics[width = \columnwidth]{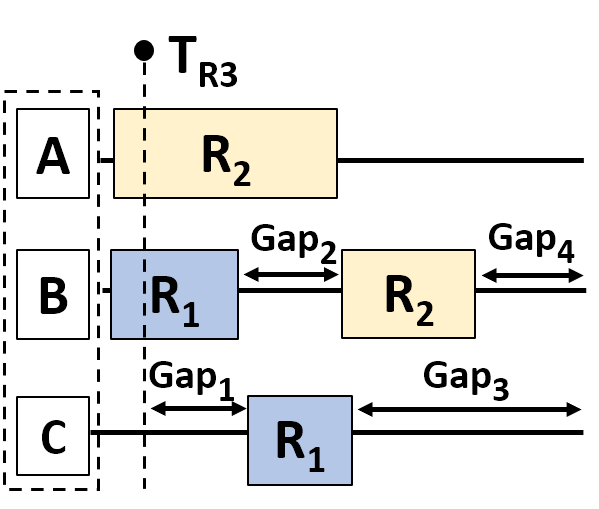}
        		\vspace*{-1.5em}
        		\caption{}
        		\label{fig:timelineBasedApproach_a}
        	\end{subfigure}
        	\begin{subfigure}[b]{0.32\columnwidth}
        		\includegraphics[width = \columnwidth]{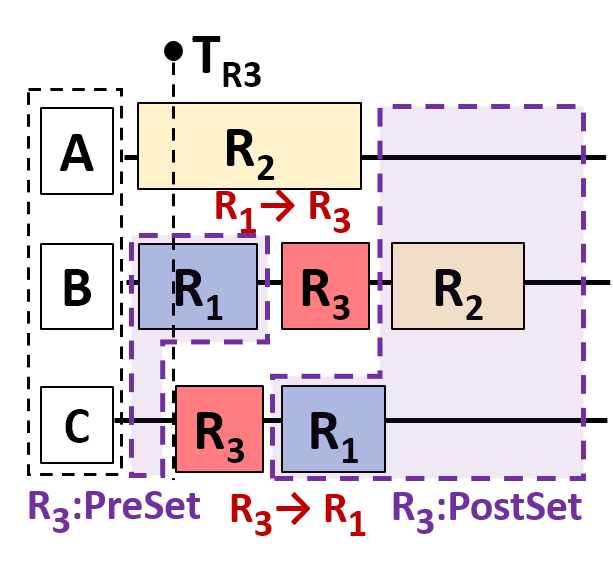}
        		\vspace*{-1.5em}
        		\caption{}
        		\label{fig:timelineBasedApproach_b}
        	\end{subfigure}
        	\begin{subfigure}[b]{0.32\columnwidth}
            	\includegraphics[width = \columnwidth]{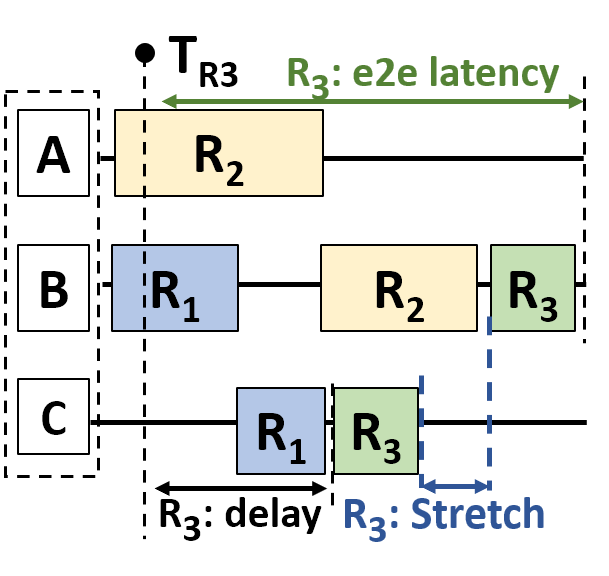}
            	\vspace*{-1.5em}
            	\caption{}
            	\label{fig:timelineBasedApproach_c}
        	\end{subfigure}
        	\vspace*{-1em}
        	\caption
        	{
                \small \it 
                {\bf Timeline Scheduler (TL) example}
                a) before scheduling $R_3$ b) trying a potential (but invalid) schedule, c) scheduling $R_3$ at the first possible gap. 
            }
            \label{fig:timelineBasedApproach}
        \end{figure}

	\noindent{\bf Timeline(TL) Scheduling: } This flexible policy uses estimates of lock-access durations, and {\it speculatively} places waiting routines into the lineage table based on these estimates. This means no routines need to wait {(for an eligibility test)} before being added to the lineage table.
	TL scheduling tries to place routines in the gaps in the lineage table without violating the lineage table invariants (Section~\ref{sec:lineage}).  An example is shown in Figures~\ref{fig:timelineBasedApproach_a},  ~\ref{fig:timelineBasedApproach_b}. Figure~\ref{fig:timelineBasedApproach_c} shows that TL  may ``stretch'' a routine's execution time due to lock waits during execution.
	To mitigate this, a new routine is delayed from starting (now) if this were to cause TL to stretch some running routine  beyond a pre-specified threshold.


    	    \begin{algorithm}[t]
        		\caption{Timeline scheduling of routine $R$}
        		\label{algo:scheduling}
        		\small
        		\begin{algorithmic}[1] 
        			
        			\Function{Schedule}{$R$, index, startTime, preSet, postSet}
        			
        			\State devID = $R[index].devID$
        			\State{duration = $lock\_access(R, devID).duration$}\label{line:duration}
        
        			\State{//return from recursion}
        			\If{{\tt $R.cmdCount$ < index}}
        			    \State{\Return true} \label{line:baseCase}
        			\EndIf 
        						
        			\State{//Find gap and pre- and post-set}
        			\State gap = getGap(devID , startTime, duration) \label{line:gap}
        			\State curPreSet = preSet $\cup$ getPreSet(lineage[devID], gap.id) 
        			\State curPostSet = postSet $\cup$ getPostSet(lineage[devID], gap.id)
        			
            		\If{{\tt curPreSet $\cap$ curPostSet = $\emptyset$} } \label{line:intersection}
            			\State{//Serialization is not violated}
            			
            			\State canSchedule = schedule($R$, index + 1, gap.startTime + duration , curPreSet, curPostSet) \label{line:deepDiveCall}
            			
            			\If{{\tt canSchedule}}
                			\State { lineage[devID].insert($R[index]$, gap)}
                			\State {\Return true}
            			\EndIf
        			\EndIf			
        			\State {//backtrack: try next gap}
        			\State{\Return {schedule($R$, index, gap.startTime + duration , preSet, postSet)} \label{line:backtrack}}
        			\EndFunction
        			
        		\end{algorithmic}
        	\end{algorithm}

        	TL scheduling uses a {\it backtrack-based search strategy} to find the best placement for a new routine in the lineage table. \Algorithm~\ref{algo:scheduling} shows the  pseudocode. 
        	We explain via an example. 
        	\Figure~\ref{fig:timelineBasedApproach_a} depicts a lock table right before routine $R_3=\{C \rightarrow B\}$ arrives at time $T_{R3}$, and has four gaps in the lineage. 
        	Starting with the first device in the routine ($C$ for $R_3$):  $\tau_{R_3}(C)$ (\Line~\ref{line:duration}), the Timeline scheduler finds the first gap in $C$'s lineage that can fit $\tau_{R_3}(C)$ (\Line~\ref{line:gap}). This is Gap 1 in \Figure~\ref{fig:timelineBasedApproach_a}.
    	    Next, the Timeline scheduler validates that this gap choice will not violate previously decided serializations. For the scheduled lock-accesses of $R_3$ so far, it builds two sets: a) \textit{preSet:} the union of all (executing and scheduled) routines  placed \textit{before} $R_3$'s lock-accesses ($\{R_1\}$ in \Figure~\ref{fig:timelineBasedApproach_b}), and b) \textit{postSet:} the union of all (executing and scheduled) routines placed \textit{after} $R_3$'s lock-accesses ($\{R_1, R_2\}$ in \Figure~\ref{fig:timelineBasedApproach_b}). The preSet and postSet of $R$ represent the routines positioned before and after $R$,  respectively, in the serialization order. { The gap choice is valid {\it if and only if} the  intersection of the preSet and the postSet is empty.} If true, the scheduler moves on to the next command of the routine. Otherwise (\Figure~\ref{fig:timelineBasedApproach_b}), the scheduler backtracks and tries the next gap (\Line~\ref{line:backtrack}). The process  repeats.

\section { {\name} Implementation}

    \begin{figure}[]
     	\centering
        \begin{subfigure}[b]{0.8\columnwidth}
            \includegraphics[width = \linewidth]{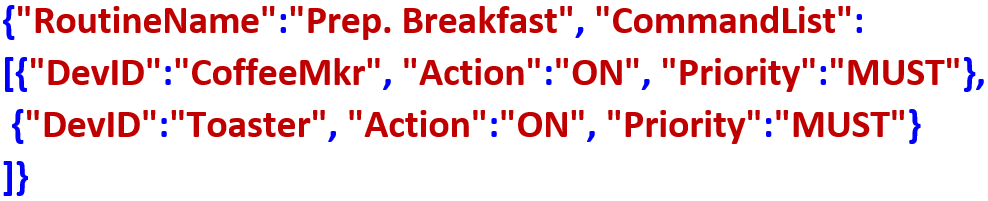}
\vspace{-0.4cm}
    		\caption{\small JSON representation of {\name} routine (part)}
    		\label{fig:safeHoomeRoutine}
        \end{subfigure}
\vspace{0.3cm}

        \begin{minipage}[b]{0.8\columnwidth}
            \centering
        	\begin{subfigure}[b]{0.49\columnwidth}
            	\includegraphics[width = \linewidth]{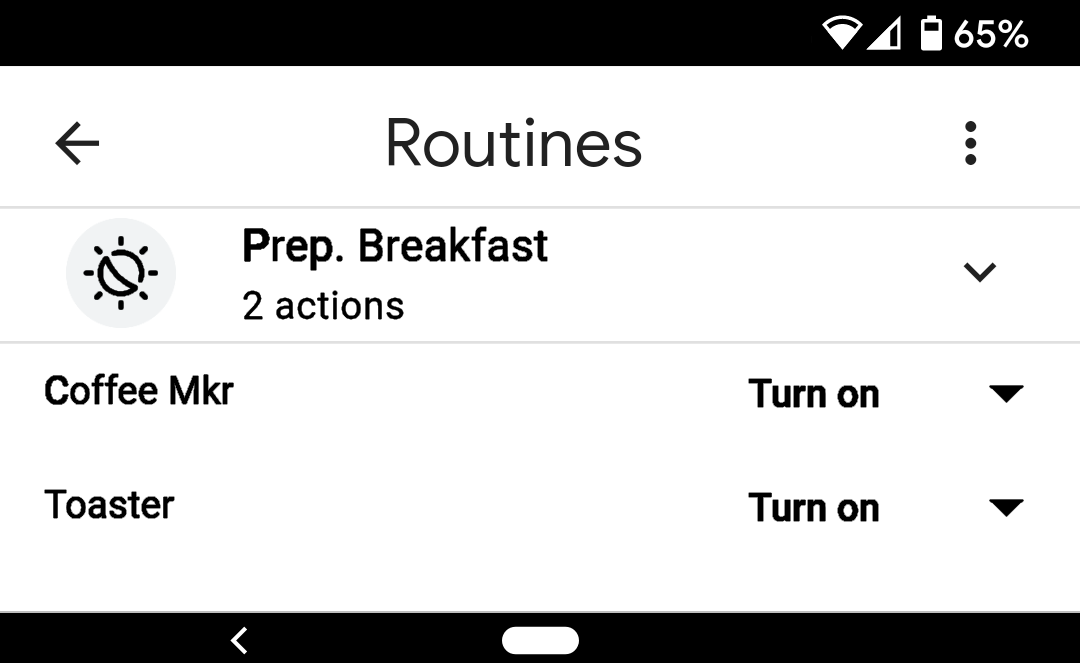}
            	\caption{\small G. Home routine~\cite{GoogleHome}}
            	\label{fig:GHomeRoutine}
        	\end{subfigure}
        	\begin{subfigure}[b]{0.49\columnwidth}
        		\includegraphics[width = \linewidth]{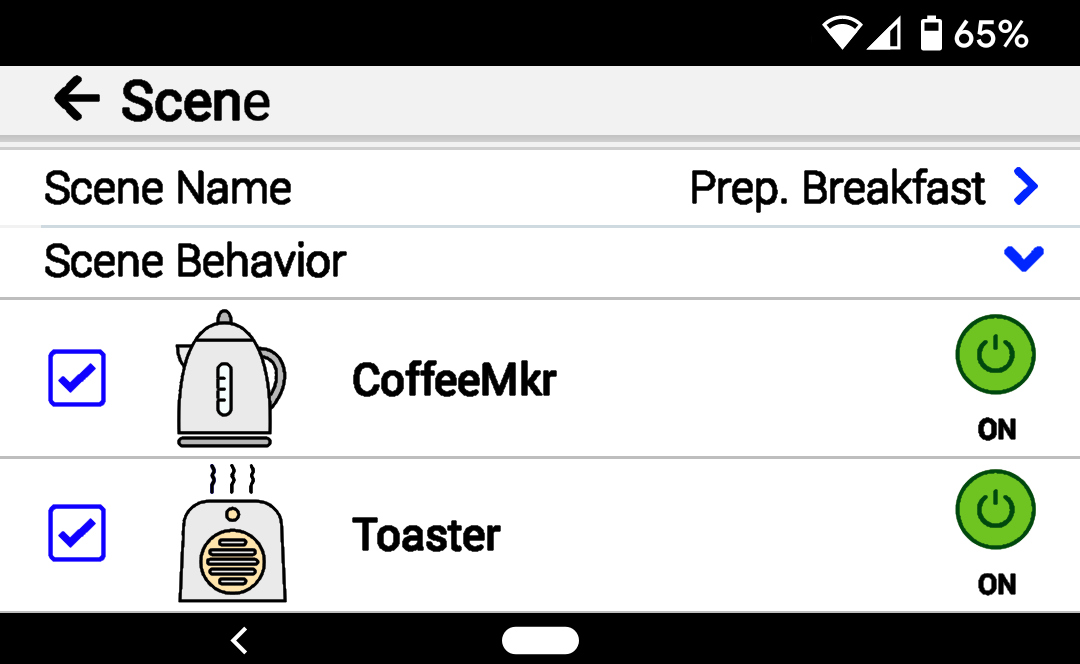}
        		\caption{\small TP-Link routine~\cite{tpLinkKasaApp}}
        		\label{fig:kasaRoutine}
        	\end{subfigure}
        \end{minipage}
    	\caption
    	{
            \small \it 
            {\bf Defining a routine ``Prepare Breakfast''}
            Two commands: i)Turn {\tt ON} Coffee Maker and ii) Turn {\tt ON} Toaster.
        }
        \label{fig:routines}
        \vspace{-1.5em}
    \end{figure}

    We implemented {\name} in 1200 core lines of Java. {\name} runs on an edge device, such as a Home Hub or an enhanced/smart access point. Our edge-first approach has two major advantages: 1) {\name} can be run in a smart home containing devices from a diverse set of vendors, and 2) {\name} is autonomous, without being affected by ISP/external network outages~\cite{XfinityOutage, ComcastOutage} or  cloud outages~\cite{AlexaOutage, GoogleHomeOutage,SmartThingsOutage}.
    
    {\name} works directly with the APIs exported by devices---commands in routines are programmed as API calls directly to devices. 
    {\name}'s routine specification is compatible with other smart home systems (\Figure~\ref{fig:routines}). Our current implementation works for TP-Link smart devices~\cite{TP-Devices, TP-DevicesInExperiment}, using the HS110Git~\cite{HS110Git} device-driver. Other devices (e.g., Wemo~\cite{Wemo}) can be supported via their device-drivers.

\begin{figure}[t]
    	\centering
    	\includegraphics[width=0.8\linewidth]{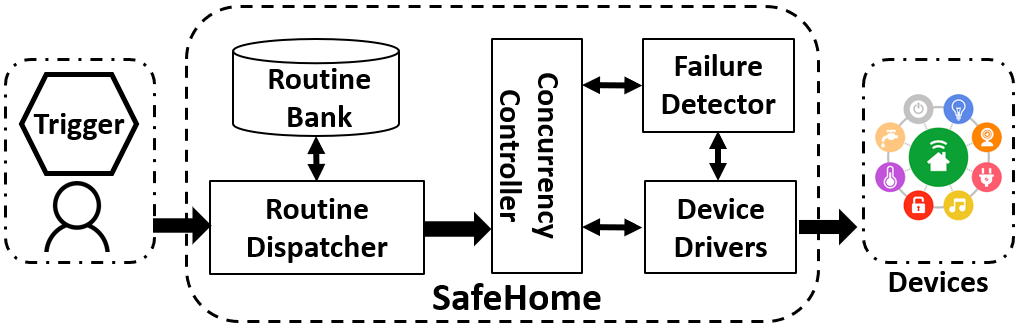}
    	\caption{\small \bf {\name} Architecture}
    	\label{fig:highLvlDesign}
    \end{figure}

    \Figure~\ref{fig:highLvlDesign} shows our implementation architecture. When a user submits routines, they are stored in the {\it Routine Bank}, from where they can be invoked either by the user or triggers, via the {\it Routine Dispatcher}. The {\it Concurrency Controller} runs the appropriate Visibility model's implementation. Apart from Eventual Visibility (\Section~\ref{sec:EVdesign}), we also implemented Global Strict Visibility (GSV), and Partitioned Strict Visibility (PSV), with  failure/restart serialization. Our Weak Visibility reflects today's laissez-faire implementation.

    The {\it Failure Detector}  explicitly checks devices by periodically (1 sec) sending ping messages. If a device does not respond within a timeout ($100$ ms by default), the failure detector marks it as {\tt failed}. We also leverage {\it implicit} failure detection by using the last heard {\name} TCP message as an implicit ack from the device, reducing the rate of pings.
        
    \section{Experimental Results}

\begin{figure*}[]
    \begin{subfigure}[b]{0.75\linewidth}
        \centering
        \captionsetup{width=.9\linewidth}
        \includegraphics[width =0.97\linewidth]{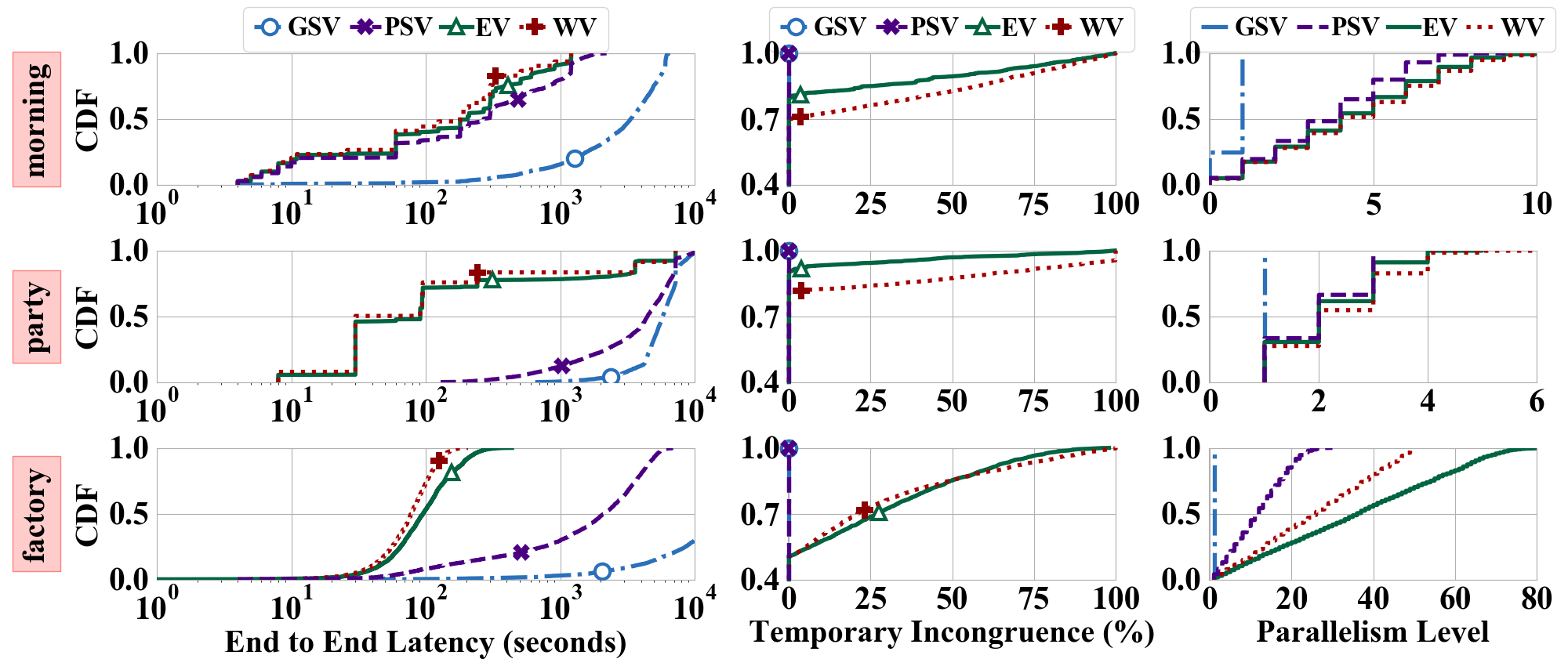}
        \caption{ {\bf Latency, Temporary Incongruence, and Parallelism for Three Scenarios.} {\it To identify lines we show one label symbol for each (plot  has many more data points). Some GSV lines may be cut to show separation between other models.  }
        }
        \label{fig:combined-real-world}
	\end{subfigure}
	\begin{subfigure}[b]{0.23\linewidth}
	    \centering
	    \captionsetup{width=.9\linewidth}
    	\includegraphics[width=0.97\linewidth]{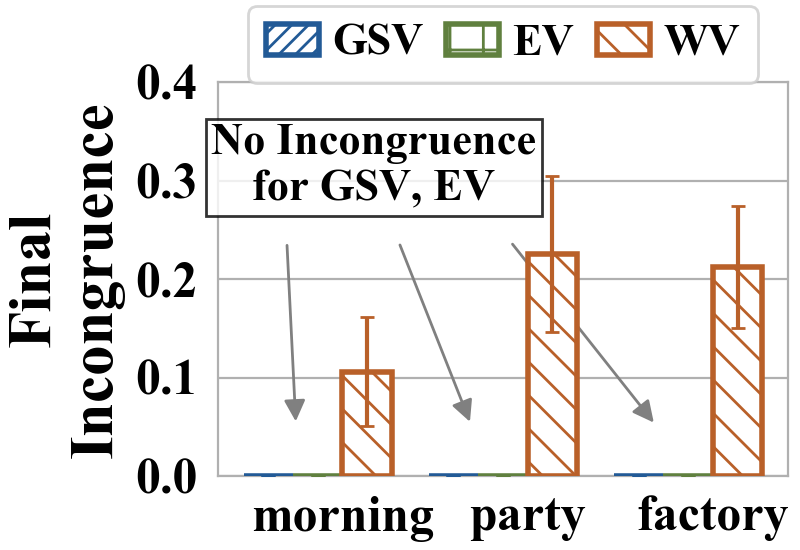}
    	\caption{\small \textbf{Final Incongruence.} \it Run with 9 routines, 100 runs per scenario, and checks if final smart home state is equivalent to some serial ordering of routines (9! possibilities). Final Incongruence measures the ratio of end states that were not congruence out of 100 runs.}
    	\label{fig:final-incongruence}
    \end{subfigure}
    \vspace*{-6pt}
    \caption{\textbf{Experiment Results with Trace-Based Scenarios}}
    \vspace*{-6pt}
\end{figure*}

    We evaluate {\name} using both workloads based on real-world deployments, and microbenchmarks. 
    The major questions we address include:
    \vspace*{-0.05cm}
    \begin{squishedEnumerate}
    
    	\item Are relaxed visibility models (like Eventual Visibility) as responsive as Weak Visibility, and as correct as Global Strict Visibility (\Section~\ref{sec:visbilitymodels})?
    	
    	\item What effect do failures have on correctness and user experience (\Section~\ref{sec:failureHandling})?
    	
    	\item Which scheduler policy (\Section~\ref{sec:EVdesign}) is the best?
    	
    	\item What is the effect of optimizations, e.g., lock leasing, commit compaction, etc. (\Section~\ref{sec:safeHome_Design})?
    	
    \end{squishedEnumerate}
\vspace*{-0.05cm}

\subsection{Experimental Setup}

        We wish to evaluate {\name} for a variety of scenarios and parameters. 
        Hence we run our implementation over an emulation, using both real-world workloads (\Section~\ref{sec:realworldexpts}) and synthetic workloads (\Section~\ref{sec:workloadDrivenEmulation} - \ref{sec:benchmark}).

    \noindent{\bf Metrics:}
        Because of the human-visible nature of {\name}, our primary evaluation metrics are also human-visible:\\
    	\noindent {\it End to end latency (or Latency):} Time between a routine's submission  and its successful completion. 
    	
        \noindent {\it Temporary Incongruence:}
            This metric measures how much the human user's actual experience  differs from a world where all routines were run serially. We take worst case behavior. Before a routine $R$ completes, if another routine $R'$ changes the state of {\it any} device  $R$ modified, we say $R$ has suffered a temporary incongruence event. The {\it Temporary Incongruence} metric measures the fraction of routines that suffer at least one such temporary incongruence event.

        \noindent {\it Final Incongruence:}
        {
            Final Incongruence measures the ratio of runs that end up in an incongruent state.
        }
        
        \noindent {\it Parallelism level:}
            This efficiency/utilization metric is the number of routines that are allowed by \name{} to execute concurrently, averaged throughout the run. To avoid domination by durations when only 0 or 1 routines run, we  only measure the metric at points when a routine starts/ends.


\subsection{Experiments with Real-World Benchmarks}
    \label{sec:realworldexpts}

    We extracted traces from three real homes (20-30 devices, multi-user families) who were using Google Home, over 2 years. We also studied two  public datasets: 1) 147 SmartThings applications~\cite{SmartApps}; and 2) IoTBench: 35 OpenHAB applications~\cite{IoTBench}. Based on these, we created three representative benchmarks: (We will make these available openly.)

    \noindent {\bf Morning Scenario:}
        This chaotic scenario has 4 family members in a 3-bed 2-bath home concurrently initiating 29 routines over 25 minutes touching 31 devices. Each user starts with a wake-up routine and ends with the leaving home routine. In between, routines cover bedroom $\&$ bathroom use, breakfast cook + eat, and sporadic routines, e.g., milk spillage cleanup. 

    \noindent {\bf Party Scenario:}
        Modeling a small party, it includes one long routine controlling the party atmosphere for the entire run, along with 11 other routines covering spontaneous events, e.g., singing time, announcements, serving food/drinks, etc.

    \noindent {\bf Factory Scenario:}
        This is an assembly line with 50 workers at 50  stages. Each stage has access to  local devices, to some devices shared with immediately preceding and succeeding stages, and to 5 global devices. 
        Each stage's routine has device access probabilities: 0.6 for local devices, 0.3 for neighbor devices, and 0.1 for global devices.
        {
            Routines are generated to keep each  worker occupied (no idle time).
        }

        {
            We trigger routines at random times while obeying  preset constraints capturing real-life logic, e.g.,  ``wake-up'' routine before ``cook breakfast'' routine. In the morning scenario, each routine occurs once per run, and for the factory scenario routines are probabilistically generated (with possible repetition). We run 1000 trials to obtain each datapoint.
        }

    \noindent {\bf Results:}
        From \Figure~\ref{fig:combined-real-world} (top row), in the morning scenario: 1) EV's latency is comparable to WV at both median and $95^{th}$ percentile, and 2) PSV has 15\% worse $90^{th}$ percentile latency than EV. Generally, the higher the parallelism level (last column), the lower the latency. For instance, EV has a median parallelism level 3$\times$ higher than GSV, and median latency 16$\times$ better than GSV. Parallelism  creates more temporary incongruences (middle column of figure). This is expected for EV. Yet, EV's (and GSV's) end state is serially equivalent while WV may end incongruently--this is shown in \Figure~\ref{fig:final-incongruence}. Thus {\it EV offers similar latencies as, but better final congruence than, WV.} Only if the user cares about temporary incongruence is PSV preferable.

        In \Figure~\ref{fig:combined-real-world} (middle row), the party scenario shows similar trends to the morning scenario with one notable exception. PSV's benefit is lower,  with only 11\% 90th percentile latency reduction from GSV (vs. 77\% in {morning}). This occurs because the single long routine blocks other routines. EV avoids this head-of-line blocking 
        because of its pre- and post-leasing.

        In \Figure~\ref{fig:combined-real-world} (bottom row), the factory scenario shows similar trends to morning scenario, except that: (i) EV's median latency is 23.1\% worse than WV, and (ii) the parallelism level is higher in EV than WV. This is due to the back-to-back arrival of multiple routines. WV executes them as-is. However, EV may delay some routines (due to device conflicts)--when the conflict lifts, all eligible routines run simultaneously, increasing our parallelism level and latency.

\subsection{Workload-Driven Emulation: Parameters}
    \label{sec:workloadDrivenEmulation}

        {
            The rest of this section performs workload-driven  experiments. Table~\ref{tab:TunableParameters} summarizes the  parameters used.
            By default we run 100 routines, 25 devices, and an average of 3 commands per routine. Each routine has a $10\%$ probability of being long-running. We run 1M trials to obtain each datapoint.
        }
    
        \begin{table}[]
        \centering
    	\resizebox{0.9\columnwidth}{!}
    	{
    	    \footnotesize
    		\begin{tabular}{|c|c|l|}
    		    \hline
    		    {\bf Name} & {\bf default} & {\bf Description}\\
    			\hline
    			$\mathcal{R}$ & $100$   &   Total number of routines\\
    			\hline
    			$\mathcal{\rho}$ & $4$ & Number of concurrent routines injected\\
    			\hline
    			$\mathcal{C}$ & $3$ & Average commands per routine (ND) \\
    			\hline
    			$\alpha$ & $0.05$ & Zipfian coefficient of device popularity\\
    			\hline
    			$\mathcal{L}_{\%}$ & $10$\% &   Percentage of long running routines\\
    			\hline
    			$|\mathcal{L}|$ & $20$ min. &   Average duration of a long running command (ND) \\
    			\hline
    			$|\mathcal{S}|$ & $10$ sec. &  Average duration of a short running command  (ND) \\
    			\hline
    			$\mathcal{M}$ & $100$\% & Percentage of ``Must'' commands of a routine\\
    			\hline
    			$\mathcal{F}$ & $0$\% &  Percentage of the failed devices\\
    			\hline
    		\end{tabular}
    	}
    	\caption{
    	    {\bf Parameterized Microbenchmark: Summary of Parameters.}
    	    {\it ND = Normal distribution.}
    	}
    	\label{tab:TunableParameters}
    	\vspace*{-6pt}
    \end{table}

\subsection{Atomicity Evaluation: Effect of Failures}
    \label{sec:failureAnalysis}
  
\begin{figure}[]
    	\centering
    	\begin{subfigure}[b]{0.49\columnwidth}
    		\includegraphics[width = \linewidth]{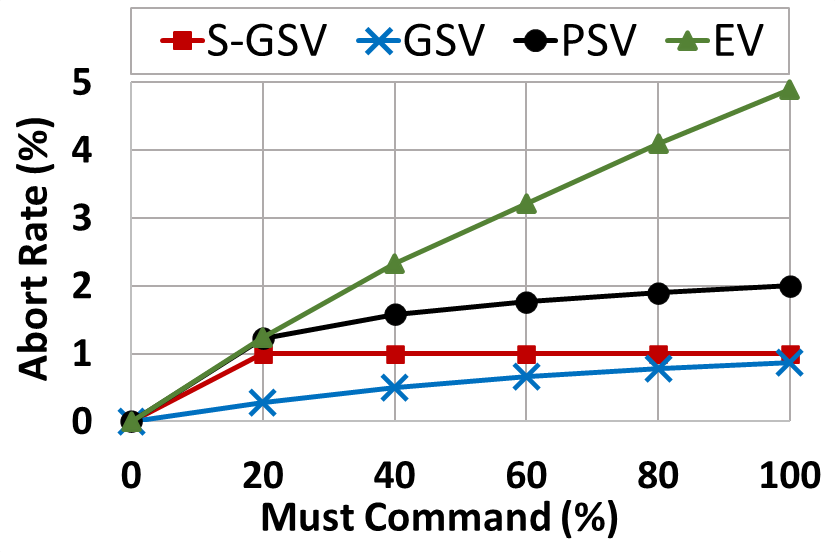}
    		\caption{Must Vs Abort Rate}
    		\label{fig:failure_MustVsAbt}
    	\end{subfigure}
    	\begin{subfigure}[b]{0.49\columnwidth}
        	\includegraphics[width = \linewidth]{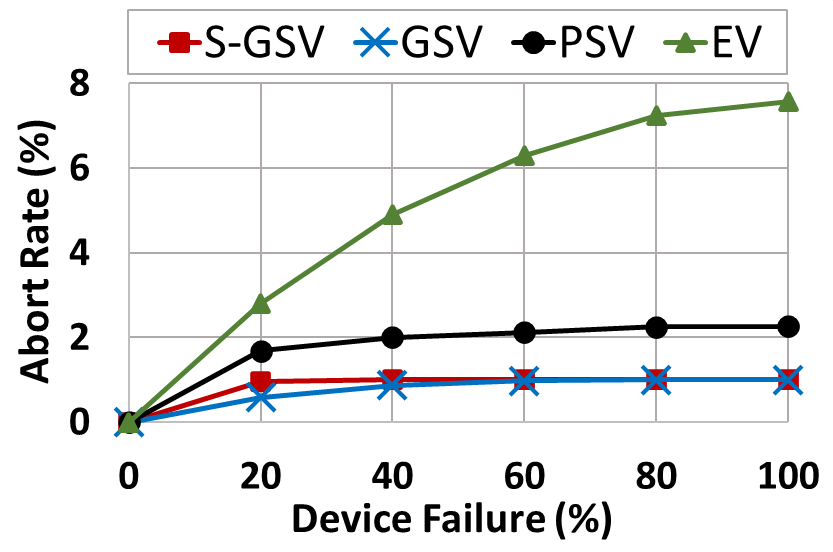}
        	\caption{Failure Vs Abort Rate}
        	\label{fig:failure_FailureVsAbt}
    	\end{subfigure}
    	
    	\begin{subfigure}[b]{0.49\columnwidth}
    		\includegraphics[width = \linewidth]{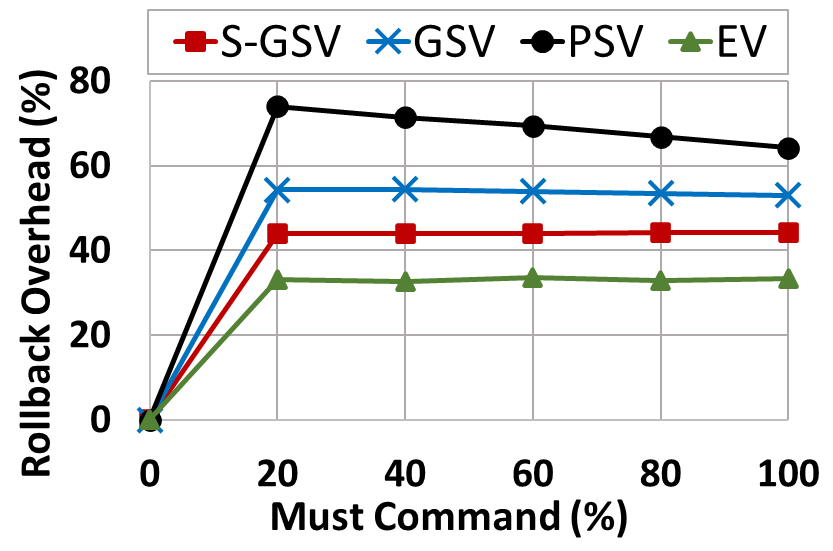}
    		\caption{Must Vs Rollback Overhead}
    		\label{fig:failure_MustVsRollback}
    	\end{subfigure}
    	\begin{subfigure}[b]{0.49\columnwidth}
        	\includegraphics[width = \linewidth]{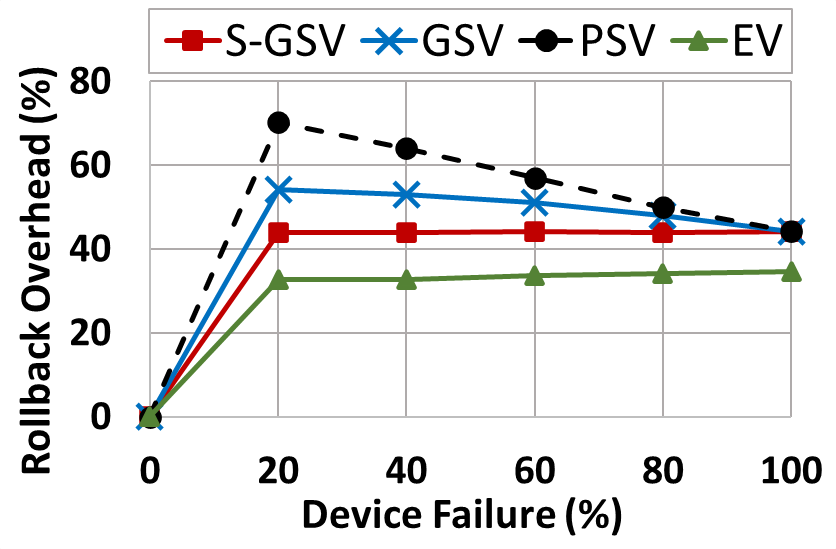}
        	\caption{Failure Vs Rollback Overhead}
        	\label{fig:failure_failureVsRollback}
    	\end{subfigure}
    	\caption
    	{
    	    {\bf Effect of Failures.}
    	    {\it Rollback Overhead = Intrusion on User. Parameters in \Table~\ref{tab:TunableParameters}.}
        }
        \label{fig:failure_and_must}
    \vspace*{-0.1cm}
    \end{figure}

    Failures abort more routines in EV because it allows high concurrency, yet EV's intrusive effect on the user (due to aborts) is the lowest of all visibility models.  \Figure~\ref{fig:failure_MustVsAbt} and \ref{fig:failure_FailureVsAbt} measure the fraction of routines aborted due to a failure. We induce fail-stop failures, where 25\% of the total devices were marked as failed at a random point during the run.
    Yet \Figure~\ref{fig:failure_MustVsRollback} and \ref{fig:failure_failureVsRollback} show that the {\it rollback overhead} of EV is smallest among all visibility models--this is the average fraction of commands rolled back, across aborted routines. PSV's  rollback overhead is higher than EV as it  aborts more at the routine's finish point (when checking up/down status of devices touched).  EV aborts affected routines earlier rather than later.     GSV and S-GSV have low abort rates because of their serial execution but have  higher rollback overheads than EV. Thus, even {\it when execution is serial, the effect of failures can be more intrusive on the human}. We conclude that EV is the least intrusive model.
    
    The plateauing in Figures~\ref{fig:failure_MustVsAbt},  \ref{fig:failure_FailureVsAbt} is due to saturation of parallelism level. The plateauing in \Figures~\ref{fig:failure_MustVsRollback},  \ref{fig:failure_failureVsRollback} is due to saturation at 
    abort-points--for GSV at 50\%, with S-GSV lower at 40\% since {\it any} device failure triggers the abort.

\subsection{Scheduling Policies}
    \label{sec:schedulingPolicies}

    \begin{figure}[]
    	\centering
    	\begin{subfigure}[b]{0.30\columnwidth}
    		\includegraphics[width = \columnwidth]{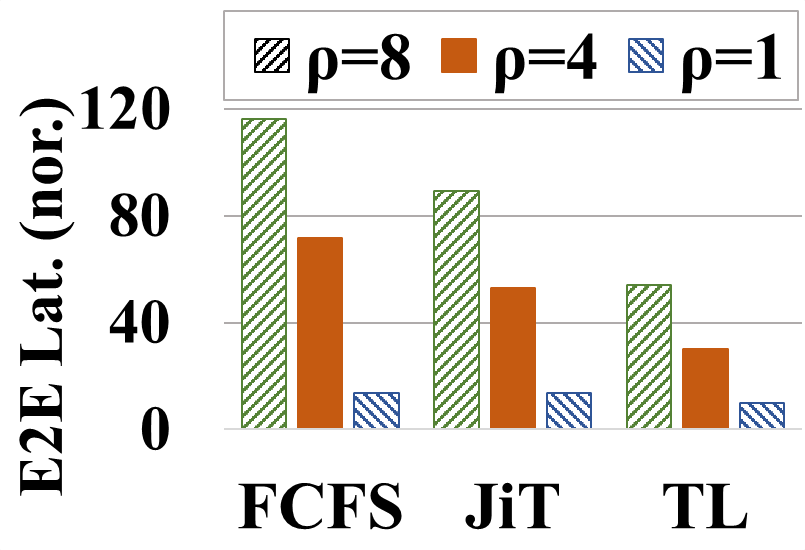}
    		\caption{ E2E Latency}
    		\label{fig:Comparison_e2eLat}
    	\end{subfigure}
    	\begin{subfigure}[b]{0.30\columnwidth}
    		\includegraphics[width = \columnwidth]{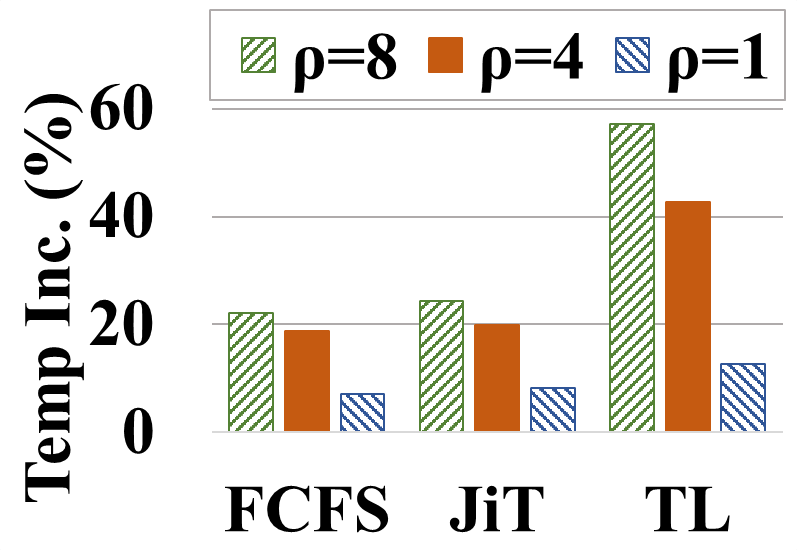}
    		\caption{Incongruence}
    		\label{fig:Comparison_Isolation}
    	\end{subfigure}
    	\begin{subfigure}[b]{0.30\columnwidth}
        	\includegraphics[width = \columnwidth]{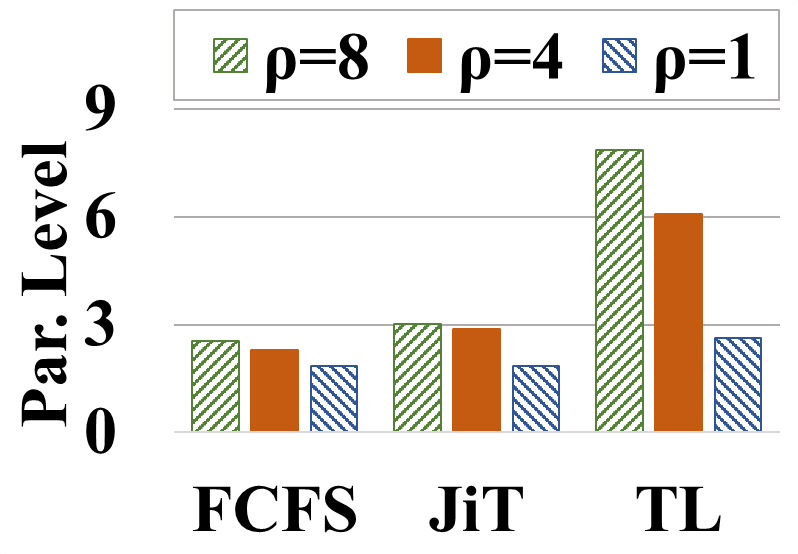}
        	\caption{ Parallelism}
        	\label{fig:Comparison_Parallelism}
    	\end{subfigure}
    	\caption
    	{
    	   { \bf Scheduling Policies.}  {\it Parameters in \Table~\ref{tab:TunableParameters}. (a) E2E Latency  normalized with routine runtime. (b) Temporary Incongruence. (c) Parallelism Level.}
        }
        \label{fig:Comparison}
    \end{figure}
    
    	

    \Figure~\ref{fig:Comparison} compares  FCFS, JiT, and Timeline (TL) scheduling policies  (\Section~\ref{sec:EVdesign}). 
    In \Figure~\ref{fig:Comparison_e2eLat} with $\rho = 4$ concurrent routines, TL is $2.36\times$ and $1.33\times$ faster than FCFS and JiT respectively.  
    {
        The benefit of TL over FCFS is due to  pre-leasing. The benefit of TL over JiT is due to opportunistic use of leasing.
    }
    TL also has  higher parallelism level (\Figure~\ref{fig:Comparison_Parallelism}) than FCFS ($2.3\times$ at $\rho = 4$) and JiT ($2.0\times$ $\rho = 4$).
    

\subsubsection{Timeline-based Eventual Visibility (TL)}
    \label{sec:Timeline_based}
 
 \begin{figure}[t]
    	\centering
    	\begin{subfigure}[]{0.49\columnwidth}
    		\includegraphics[width = \linewidth]{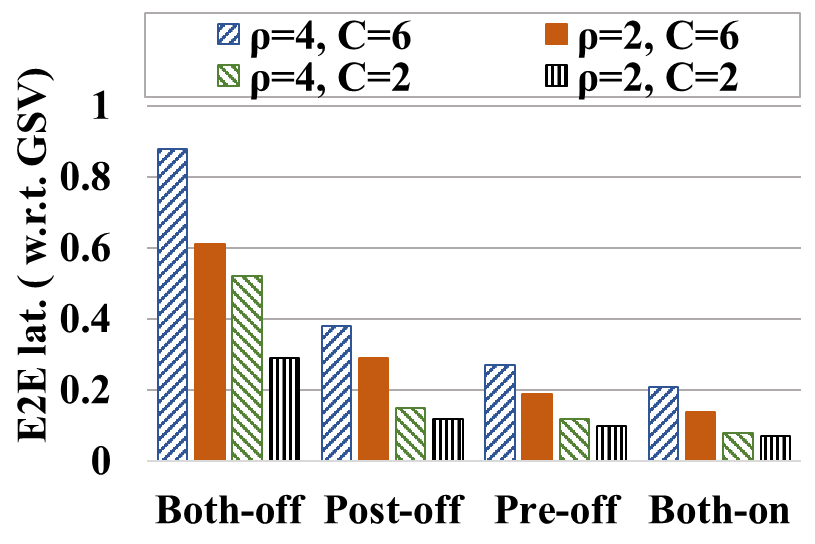}
    		\caption{Normalized E2E Latency}
    		\label{fig:analyzeEV_latencyOverhead}
    	\end{subfigure}
    	\begin{subfigure}[]{0.49\columnwidth}
    		\includegraphics[width = \linewidth]{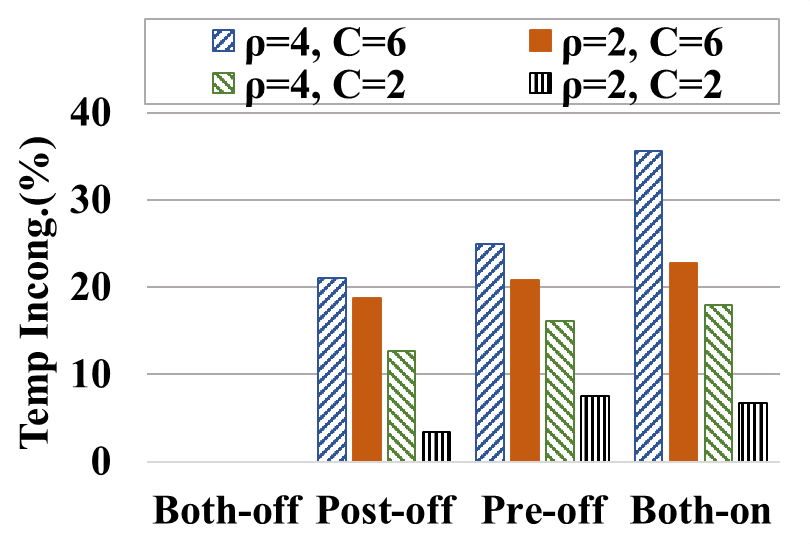}
    		\caption{Temporary Incongruence (\%)}
    		\label{fig:analyzeEV_Isolation}
    	\end{subfigure}
    	
    	\begin{subfigure}[]{0.49\columnwidth}
        	\includegraphics[width = \linewidth]{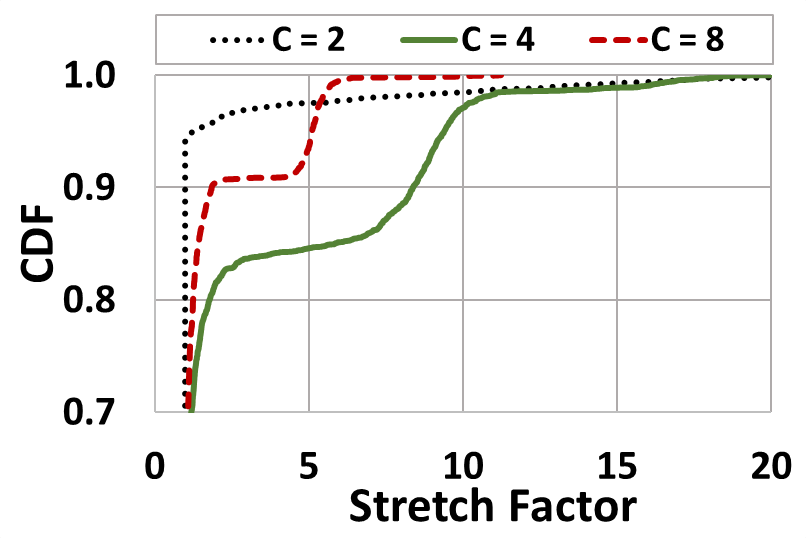}
        	\caption{CDF of Stretch Factor}
        	\label{fig:analyzeEV_Stretch_commandCount}
    	\end{subfigure}
    	\begin{subfigure}[]{0.49\columnwidth}
        	\includegraphics[width = \linewidth]{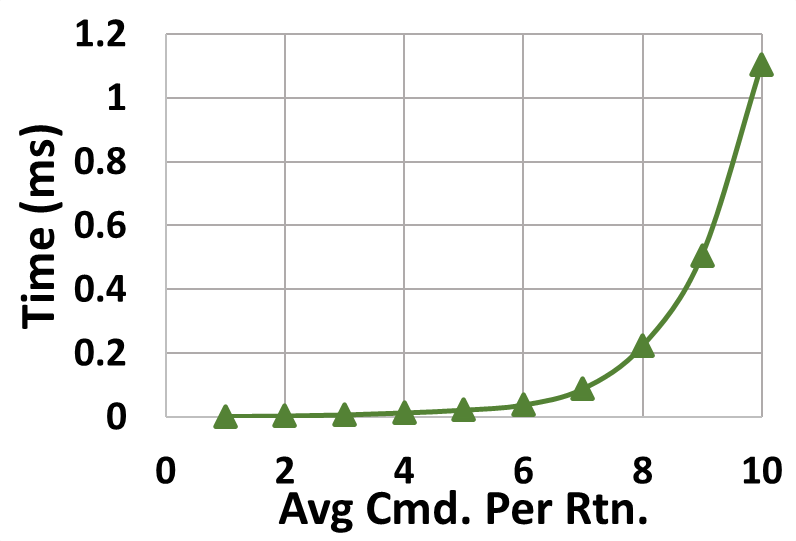}
        	\caption{\Algorithm~\ref{algo:scheduling} Insertion Time}
        	\label{fig:analyzeEV_Stretch_rtnInsertionTime}
    	\end{subfigure}
\vspace*{-0.2cm}
    	\caption
    	{
    	    {\bf TL Scheduler under EV.} {\it Parameters in Table~\ref{tab:TunableParameters}.} 
    	}
    \vspace*{-0.2cm}
    \end{figure}

     \Figure~\ref{fig:analyzeEV_latencyOverhead} and \ref{fig:analyzeEV_Isolation} show that disabling leasing reduces temporary incongruence but significantly increases  latency. Turning off {\it both} pre and post leasing increases latency (from Both-on to Both-off) by between $3\times$ to $5.5\times$ (as concurrency level $\rho$ and commands per routine $\mathcal{C}$ are varied).
    {Post-leases are more effective than pre-leases: disabling the former raises latency by between $71\%$ to $107\%$, while disabling the latter raises latency from between $29\%$ to $50\%$. Post-leasing opportunities are more frequent than pre-leasing ones because the former does not require changing the serialization order (the latter does). 
    }
    These trends are true for all combinations of $\rho,\mathcal{C}$.

    TL might also ``stretch'' routines (\Figure~\ref{fig:timelineBasedApproach_c}). \Figure~\ref{fig:analyzeEV_Stretch_commandCount} shows  {\it stretch factor}, measured as the time between a routine's actual start  (not submission) and actual finish, divided by the ideal (minimum) time to run the routine. With routine size, stretch factor rises at first (at $\mathcal{C} = 2$  only 5\% routines have stretch $>1$, vs. 25\% at $\mathcal{C} = 4$) but then drops ($15\%$ at $\mathcal{C} = 8$). Essentially the lock-table saturates beyond a $\mathcal{C}$, creating fewer gaps and forcing EV to append new routines to the schedule.

    We used a  Raspberry Pi 3 B+~\cite{pi3B} to run TL as the home hub (15 devices, 30 routines). \Figure~\ref{fig:analyzeEV_Stretch_rtnInsertionTime} shows it takes only 1 ms to schedule a large routine with  10 commands. Surveys show typical routines today contain 5 commands or fewer~\cite{SmartApps, IoTBench}, hence our scheduler is fast in practice.

\subsection{Parameterized Microbenchmark Experiments}
    \label{sec:benchmark}

    {
    }
    
    \begin{figure}[]
    	\centering
    	\begin{subfigure}[b]{0.49\columnwidth}
    		\includegraphics[width = \columnwidth]{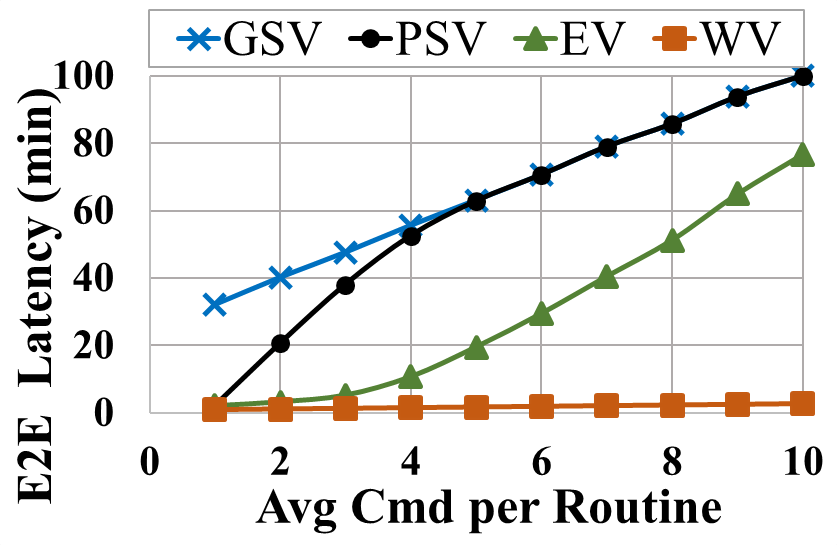}
    		\caption{End to End latency}
    		\label{fig:E2EVsCmd}
    	\end{subfigure}
    	\begin{subfigure}[b]{0.49\columnwidth}
        	\includegraphics[width = \columnwidth]{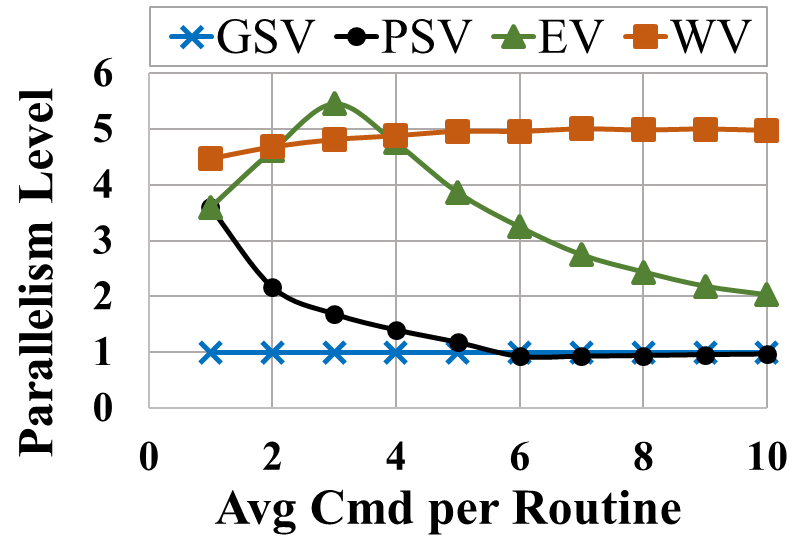}
        	\caption{Parallelism level (\%)}
        	\label{fig:parallelismVsCmd}
    	\end{subfigure}
    	
    	\begin{subfigure}[b]{0.49\columnwidth}
    		\includegraphics[width = \columnwidth]{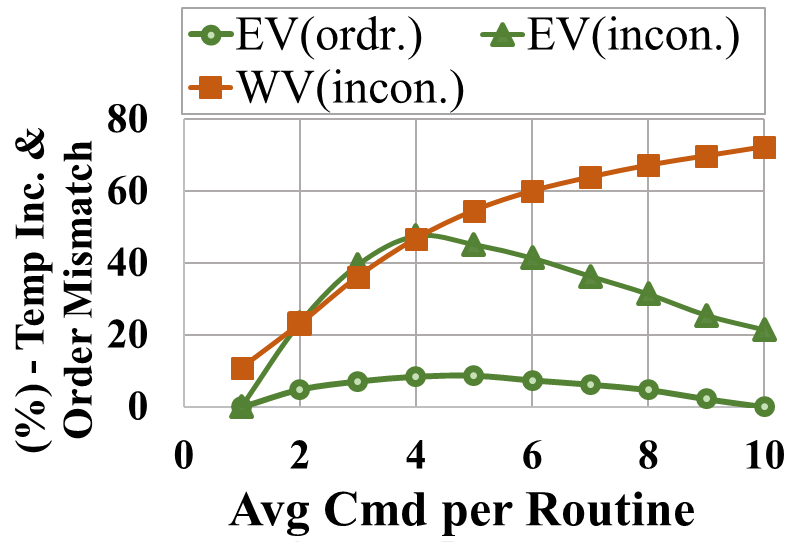}
    		\caption{Temporary Incongruence \& Order Mismatch (\%)}
    		\label{fig:isolation_orderMismatch_VsCmd}
    	\end{subfigure}
    	\begin{subfigure}[b]{0.49\columnwidth}
    		\includegraphics[width = \columnwidth]{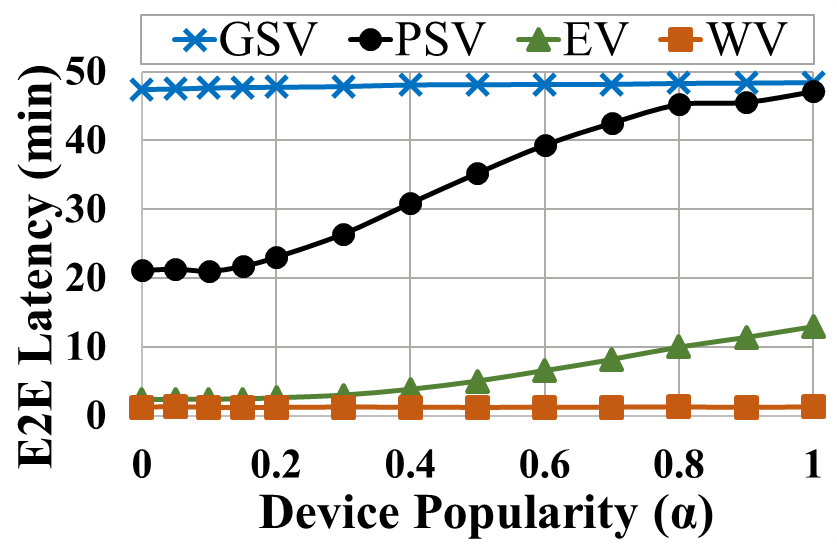}
    		\caption{Device Popularity vs. Latency}
    		\label{fig:E2EVsZipf}
    	\end{subfigure}
        \vspace*{-0.25cm}  
    	\caption
    	{
    	    {\bf Impact of Routine size ($\mathcal{C}$) and device popularity ($\alpha$).}
    	    {\it PSV and GSV are always zero and omitted in (c).}
        }
    \end{figure}
    
     \noindent{\bf Commands per routine $(\mathcal{C})$:}
        \Figure~\ref{fig:E2EVsCmd}, \ref{fig:parallelismVsCmd} show GSV's latency rises as routines contain more commands. With smaller  routines, PSV  is close to EV and WV, but as routines contain more commands, PSV quickly approaches GSV. While EV has a similar trend, it stays  faster than GSV and PSV. Parallelism level and temporary incongruence follow this trend. Finally, EV's peaking behavior and eventual convergence towards GSV (\Figure~\ref{fig:isolation_orderMismatch_VsCmd})
        occur since beyond a certain routine size ($\mathcal{C}$=4), pre/post-leasing opportunities decrease.
        
    \noindent{\bf Device popularity $(\alpha)$:}
        Using a Zipf distribution for device access by routines,  \Figure~\ref{fig:E2EVsZipf} shows that increasing $\alpha$ (popularity skew) causes EV's latency to stay close to WV. More conflict slows PSV quickly down to GSV. 
        
\begin{figure}[]
    	\centering
    	\begin{subfigure}[b]{0.49\columnwidth}
        	\includegraphics[width = \linewidth]{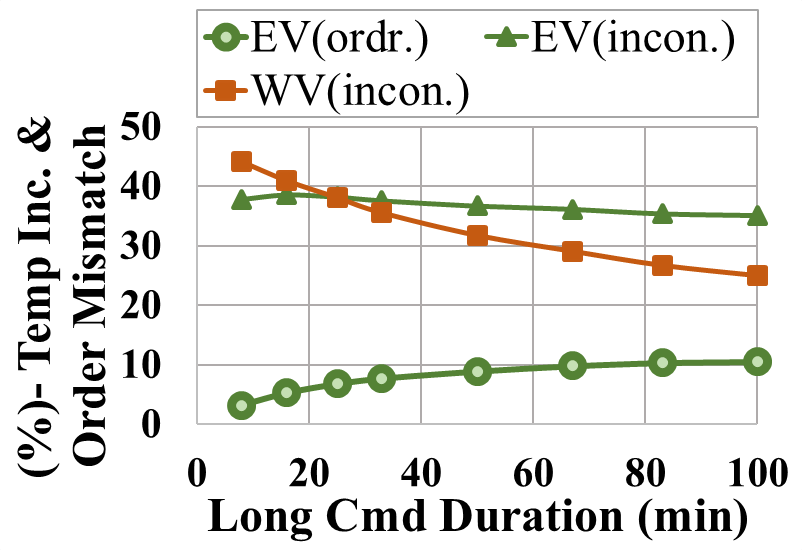}
        	\caption{}
        	\label{fig:isolationViolation_orderMismatch_LRduration}
    	\end{subfigure}
    	\begin{subfigure}[b]{0.49\columnwidth}
        	\includegraphics[width = \linewidth]{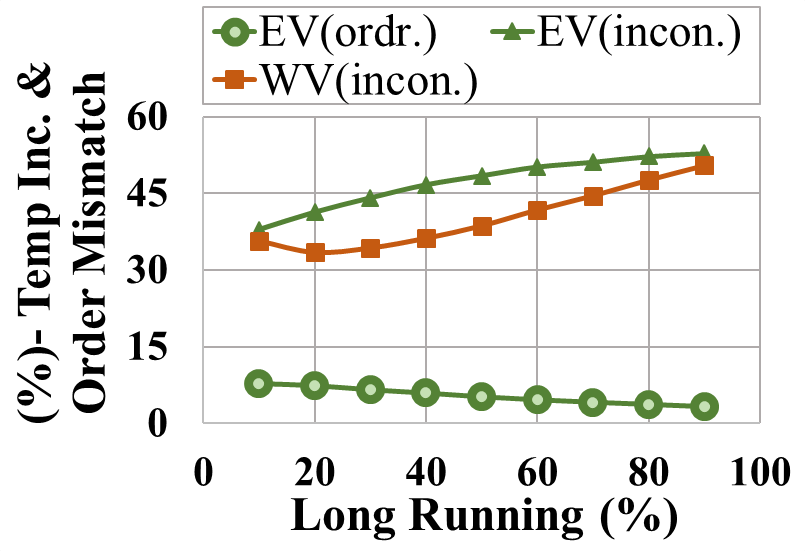}
        	\caption{}
        	\label{fig:isolationViolation_orderMismatch_LRpercent}
        \vspace{-0.5em}    
    \end{subfigure}
     	\caption
    	{
    	    \bf Impact of: (a) long routine duration ($|\mathcal{L}|$), and (b) percentage of long running routines  ($\mathcal{L}_{\%}$).
        }
        \label{fig:LRduration_and_LRpercent}
    \end{figure}

    \noindent{\bf Long running routines: }
        As the long running routine length $|\mathcal{L}|$ rises (\Figure~\ref{fig:isolationViolation_orderMismatch_LRduration}),
        temporary incongruences decrease since the run is now longer, routines are spread temporally, and less likely to conflict. Increasing the number of long running routines ($\mathcal{L}_{\%}$) increases the chance of conflict, causing more temporary incongruence. (\Figure~\ref{fig:isolationViolation_orderMismatch_LRpercent}). The {\it order mismatch}---how much the final serialization order differs from the submission order of routines, using swap distance:
        i) rises as routines get longer   (\Figure~\ref{fig:isolationViolation_orderMismatch_LRduration}), ii) but  falls with as more routines are longer  (\Figure~\ref{fig:isolationViolation_orderMismatch_LRpercent}), because post-leases dominate. Overall,  order mismatch stays low, between 3\%-10\%.
        
    \section{Related Work}

    \noindent{\bf Support for Routines:}
        Routines are supported by Alexa~\cite{Alexa}, Google Home~\cite{GoogleHome}, and others~\cite{routine:notReliable1, routine:notReliable2, routine:notReliable3}. iRobot's Imprint~\cite{iRobot:Imprint_whatIs, iRobot:Imprint_gettingStrtd} supports  long-running routines,  coordinating between a vacuum~\cite{iRobot:roomba} and a mop~\cite{iRobot:brava}. All these systems only support best-effort execution (akin to WV).

    \noindent{\bf Consistency in Smart Homes:} 
        {\name} can be used orthogonally with either: i)  transactuations~\cite{Transactuations}, which provides a consistent soft-state, {or} ii) APEX~\cite{APEX},  {which} ensures safety by automatically discovering and executing prerequisite commands. These two systems maintain strict isolation by sequentially executing conflicting routines, making them both somewhat akin to PSV.

    \noindent {\bf Abstractions:}
        IFTTT~\cite{IFTTT} represents the home as a set of simple conditional statements, while HomeOS~\cite{HomeOS} provides a PC-like abstraction for the home where devices are analogous to peripherals in a PC. Beam~\cite{Beam} optimizes  resource utilization by partitioning  applications across devices. 
        These and other abstractions for smart homes~\cite{OpenHAB,Zapier,MicrosoftFlow,AutomateIO, Workflow}  do not address failures or concurrency.
    
    \noindent {\bf Concurrency Control:}
        Concurrency control is well-studied in databases~\cite{ConcurrencyCtrl1}. Smart Home OSs like HomeOS, SIFT, and others~\cite{HomeOS, SIFT, DependencyManagement, DepSys} explore different concurrency control schemes. However, none of these explore visibility models. Classical task graph scheduling algorithms~\cite{taskGraph1, taskGraph2, taskGraph3, taskGraph4, taskGraph5, SCT, ETF} do not tackle \name's specific scheduling problem. 

    \noindent \textbf{ACID Properties applied in Other Domains:} There is a rich history of leveraging transaction-like ACID properties in many domains. Examples  include work in  software-defined networks
    to guarantee update consistency~\cite{sdn-acid1, sdn-acid2} and for  robustness~\cite{sdn-acid-robust}. ACID has also been applied in transactional memory \cite{transactional-memory-book, MetaTM, transactional-consistency1, transactional-consistency2} and pervasive computing~\cite{ecampus}.  
    \section{Conclusion}

    {\name} is: i) the first implementation of relaxed visibility models for smart homes running concurrent routines, and ii) the first system that reasons about failures alongside concurrent routines. We find that:\\
    %
        (1) Eventual Visibility (EV) provides the best of both worlds, with: a) user-facing responsiveness (latency) only  $0\%- 23.1\%$ worse than today's Weak Visibility (WV), and b) end state congruence  identical to the strongest model Global Strict Visibility (GSV).  \\ 
        (2) When routines abort due to failures, EV rolls back the fewest commands among all models. \\
        (3) Lock leasing improves latency by $3\times-5.5 \times$. \\
        (4) Compared to competing policies (FCFS and JiT), Timeline Scheduling improves  { latency} by $1.33\times - 2.36\times$ and parallelism by $2.0\times - 2.3\times$.

    \balance
    {
        \bibliographystyle{plain}
        \bibliography{resources/ref}
    }
    
\end{document}